\documentclass[12pt]{article}

\pdfoutput=1

\usepackage[english]{babel}
\usepackage{fancyhdr}
\usepackage{amsmath}
\usepackage{amssymb}
\usepackage{amsfonts}
\usepackage{psfrag}
\usepackage{cite}
\usepackage[applemac]{inputenc}
\usepackage{graphicx,wrapfig}
\usepackage[bf,footnotesize]{caption2}
\usepackage{pstricks}

\newcommand{\beq}{\begin{equation}}
\newcommand{\eeq}{\end{equation}}
\newcommand{\bea}{\begin{eqnarray}}
\newcommand{\eea}{\end{eqnarray}}
\newcommand{\bag}{\begin{align}}
\newcommand{\eag}{\end{align}}

\newcommand{\GeV}{\,\mathrm{GeV}}
\newcommand{\TeV}{\,\mathrm{TeV}}

\newcommand{\fb}{\,\mathrm{fb}}

\newcommand{\eq}[1]{Eq.~(\ref{#1})}

\newcommand{\vev}[1]{\langle {#1} \rangle}

\newcommand{\nn}{\nonumber}

\newcommand{\SU}{\textrm{SU}}
\newcommand{\SO}{\textrm{SO}}
\newcommand{\U}{\textrm{U}}
\newcommand{\Sp}{\textrm{Sp}}

\newcommand{\Hsym}{\mathcal{H}}
\newcommand{\Gsym}{\mathcal{G}}

\begin{document}

\baselineskip=18pt

\setcounter{footnote}{0}
\setcounter{figure}{0}
\setcounter{table}{0}


\begin{titlepage}

\begin{center}
\vspace{1cm}

{\huge \bf  
Composite Higgses}

\vspace{.8cm}

{\bf Brando Bellazzini$^{a,b}$, Csaba Cs\'aki$^c$ and Javi Serra$^c$}

\vspace{.5cm}

$^a$ {\it Institut de Physique Th\'eorique, CEA-Saclay and CNRS URA-2306\\
 F-91191 Gif-sur-Yvette Cedex, France}

\vspace*{0.1cm}

$^b$ {\it Dipartimento di Fisica e Astronomia, Universit\`a di Padova and INFN Sezione di Padova\\ Via Marzolo 8, I-35131 Padova, Italy}

\vspace*{0.1cm}

$^c$ {\it  Department of Physics, LEPP, Cornell University, Ithaca, NY 14853, USA}

\end{center}
\vspace{.8cm}

\begin{abstract}
\medskip
\noindent
We present an overview of composite Higgs models in light of the discovery of the Higgs boson. The small value of the physical Higgs mass suggests that the Higgs quartic is likely loop generated, thus models with tree-level quartics will generically be more tuned. We classify the various models (including bona fide composite Higgs, little Higgs, holographic composite Higgs, twin Higgs and dilatonic Higgs) based on their predictions for the Higgs potential, review the basic ingredients of each of them, and quantify the amount of tuning needed, which is not negligible in any model. We explain the main ideas for generating flavor structure and the main mechanisms for protecting against large flavor violating effects, and present a summary of the various coset models that can result in realistic pseudo-Goldstone Higgses. We review the current experimental status of such models by discussing the electroweak precision, flavor and direct search bounds, and comment on UV completions and on ways to incorporate dark matter. 
\end{abstract}

\bigskip

\end{titlepage}


\section{Introduction}
\setcounter{equation}{0}
\setcounter{footnote}{0}

The discovery of the Higgs boson~\cite{higgsdiscovery} with mass $m_h \approx 125$ GeV has been an important milestone in particle physics. It allows us for the first time to finally completely fix the parameters of the SM Higgs potential  
\begin{equation}
\label{SMpot}
V(h) = -\mu^2|H|^2 + \lambda |H|^4 \ ,
\end{equation}
where $\langle H\rangle = v/\sqrt{2}$, $v= 246$ GeV. The resulting experimental values are
\begin{equation}
\label{exp}
\mu^2_{exp} \approx (89 \GeV)^2 \ , \quad \lambda_{exp} \approx 0.13 \ .
\end{equation}
It also started to seriously weed out and constrain the once-crowded arena of models of electroweak symmetry breaking and the TeV scale: plain Technicolor/Higgsless~\cite{technicolor,higgsless} models are excluded, while the simplest supersymmetric models have a difficult time reproducing the observed value of the Higgs mass. The absence of observation of missing energy events puts strong lower limits on masses of superpartners. This review focuses on the other viable option: natural electroweak symmetry breaking from strong dynamics, where the strong dynamics produces a light composite Higgs doublet. 

The idea of a composite Higgs boson goes back to Georgi and Kaplan in the 80's~\cite{oldcompohiggs}, where it was also recognized that making it a Goldstone boson could also render the Higgs lighter than the generic scale of composites. The idea of composite Higgses has re-emerged in the guise of warped extra dimensional models in the late 90's~\cite{RS1, ADMS, HewettPomarol}, and then in the form of little Higgs models~\cite{ArkaniHamed:2002qx,ArkaniHamed:2002qy} in the early 2000's, when the crucial ingredient of collective breaking was added. Later it was realized that the pseudo-Goldstone idea with collective breaking is very easily realized in extra dimensional models, where the Higgs is identified with a component of the gauge field, giving rise to the holographic composite Higgs models~\cite{Contino:2003ve,Agashe:2004rs}, building on earlier important work~\cite{Hosotani,ABQ,earlygaugehiggsrefs}. The generic features of these constructions have been condensed into a simple 4D effective description~\cite{Giudice:2007fh,Barbieri:2007bh}. 

This review aims at explaining the main ideas behind the various types of composite Higgs scenarios, to contrast their main features, critically compare them and present the main experimental constraints on them. There are several other nice reviews on this topic with somewhat different points of view or emphasis: for a detailed description of the pseudo-Goldstone composite Higgs models see~\cite{Contino:2010rs}, for little Higgs models see~\cite{Schmaltz:2005ky,Perelstein:2005ka}, for extra dimensions and holography see~\cite{CsakiTASI02, Csaki:2005vy,Gherghetta:2006ha,Sundrum:2005jf,Serone:2009kf,Rattazzi:2003ea,Ponton:2012bi,TonyTASI}, while a recent summary of non-standard Higgs models can be found in~\cite{alternatives}. We will not follow the historical order of developments: instead we will present everything from the point of view of a 4D low-energy effective theory. 

We start by explaining the consequences of the recent measurement of the value of the Higgs mass on the parameters of the Higgs potential: both the mass and quartic self coupling are independently fixed. A light Higgs mass of 125 GeV implies a small quartic, which is more likely to point toward a loop induced quartic rather than a tree-level one. We present a simple parametrization of the potential suitable for pseudo-Goldstone composite Higgs models, and the tuning necessary to obtain this potential. In Sec.~\ref{classification} we classify the various types of composite Higgs models based on their predictions for the Higgs potential and quantify the expected amount of tuning in these models. Sec.~\ref{yukawa} contains the discussion of the various possible mechanisms for generating the Yukawa couplings, and for protecting from large flavor changing effects. We review the various coset models that can give rise to realistic patterns of symmetry breaking with SM-like Higgs bosons in Sec.~\ref{cosets}. The signals and constraints on the composite Higgs models are summarized in Sec.~\ref{signals}, and we finally comment on UV completions in Sec.~\ref{sec:UVcompletions}.

\section{The Higgs Potential of Composite Higgs Models and Tuning\label{potential}}
\setcounter{equation}{0}
\setcounter{footnote}{0}

One can nicely classify the various types of composite Higgs models by the size of the Higgs potential and also by the mechanism that generates the Yukawa couplings, in particular  for the top quark. We will first focus on the generic features of the potential, in order to categorize in Sec.~\ref{classification} composite Higgs models based on the particulars of such a potential, and finally in Sec.~\ref{yukawa} we will discuss the various mechanisms for the generation of the Yukawa couplings.\\

While the numerical values of the parameters in the Higgs potential (\ref{SMpot}) are now fixed, there are several different dynamical ways in which one can arrive at this potential. We will make the following assumptions regarding the dynamics responsible for generating the potential:

\begin{itemize}
\item The Higgs is a composite with a scale of compositeness given by $f$. 
\item There is a hierarchy between the Higgs VEV $v$ and the scale $f$: $v/f< 1$ such that the Higgs potential can be expanded in powers of $h/f$.\footnote{
In most cases this is not even necessary, given that the leading contributions to the potential can be arranged into only two definite functions of $h/f$.}
\item The Higgs potential is (fully or partially) radiatively generated. This is generically the case when the Higgs is also a pseudo-Goldstone boson (pGB). We will also assume that the potential vanishes in the limit when the SM couplings vanish. 
\end{itemize}

Using these assumptions the leading terms in the Higgs potential can be parameterized by (using $h=\sqrt{2} H$)
\begin{equation}
\label{pot}
\Delta V(h) = \frac{g_{SM}^2 \Lambda^2}{16 \pi^2} \left( -a |h|^2 + b \frac{|h|^4}{2f^2} \right) \ ,
\end{equation}
where $g_{SM}$ is a typical SM coupling, the largest of which corresponds to the top Yukawa $g_{SM}^2 \sim N_c y_t^2$. We have also introduced the scale $\Lambda$, that sets the overall size of the potential. Typically, this will be given by the mass of the state that is responsible for cutting off the quadratic divergence of the Higgs, so generically $\Lambda \sim m_*$. 
To fit the observed Higgs VEV and mass, the parameters $a$, $b$, $f$ and $\Lambda$ have to satisfy 
\begin{equation}
\label{exp}
(246 \GeV)^2 = v^2 = \frac{a}{b}  f^2 \ , \quad (125 \GeV)^2 = m_h^2 = 4 \, b \, v^2 \frac{g_{SM}^2}{16\pi^2} \frac{\Lambda^2}{f^2}  \ .
\end{equation}
We can then classify a composite Higgs model by the magnitudes of the parameters $\Lambda$, $a$, and $b$. Before we do so, we would like to make some important general remarks regarding the perturbative nature of the physics responsible for the Higgs potential and the consequences of this for fine-tuning. 

One of the main physical consequence of the magnitude of the recently measured Higgs mass is that the physics generating the Higgs potential should be weakly coupled. The experimental value of the quartic  is $\lambda_{exp} \approx 0.13$, which is of the order expected for a weakly coupled one-loop diagram. The loop factor $L$ is given by
\begin{equation}
\label{loop}
L = 2 \frac{g_{SM}^2 (\Lambda^2/f^2)}{16\pi^2} \sim 0.15 \left(\frac{g_{SM}}{\sqrt{N_c} y_t}\right)^2
\left(\frac{\Lambda/f}{2}\right)^2 \ ,
\end{equation}
where the separation between $\Lambda \sim m_*$ and $f$ determines the magnitude of the coupling of the states at $m_*$, $g_* = \Lambda /f$. We can see that for $g_* \sim 2$ the loop is about the right size for the value of the observed quartic. 
This lets us to conclude that the new physics responsible for cutting off the potential is weakly coupled, 
\begin{equation}
\label{weakk}
g_{*} \equiv \Lambda/f \ll 4\pi \ ,
\end{equation}
implying that the mass scale for new particles appears much before the true strong coupling scale $\Lambda_C \sim 4 \pi f$ is reached. While this perturbativity sounds like a welcome news for the calculability of the Higgs potential, it is also the origin of the tuning for these composite Higgs models. If the idea of a true loop-induced potential with a loop factor $L\sim 0.15$ is taken seriously, one would also expect the same factor to set the magnitude of the Higgs mass parameter, yielding the relation $f^2= \mu^2/L \approx v^2$.
However, as we will see in Sec.~\ref{ewpt} and \ref{higgscouplings}, electroweak precision tests (EWPT's) and the Higgs coupling measurements imply that $f > v$, leading to a tension with the expectation from a generic weakly coupled loop induced Higgs potential. This tension is the origin of the fine-tuning in these models: a fully natural loop-induced Higgs potential would require $f\sim v$, while EWPT and Higgs couplings require $f > v$. In practice the tuning required to get around this tension is to have several contributions to $a$ and $b$ (along with their associated $g_{SM}^2$ and $\Lambda^2$), which will then partially cancel to give an effective $a/b < 1$. Note, that lowering the coupling $g_*$ is actually not a possibility for finding non-tuned Higgs potentials with larger $f$'s: while formally the relation 
\begin{equation}
\mu^2 = 2 a \frac{g_{SM}^2 g_*^2}{16\pi^2} f^2
\end{equation}
can be satisfied for $f > v$ if $g_*$ is lowered, we actually know that $g_* f$ is a physical mass scale where new particles appear, and can thus not be too low experimentally. Also, in most models $g_{SM}$ is a derived quantity (from couplings of several BSM states related to $g_*$) usually implying relations of the form $g_{SM}<g_*$, which also sets a lower bound on how small $g_*$ can be. Finally, taking $g_* < g_{SM}$ would run counter to the philosophy of composite Higgs models, where a strongly interacting sector is expected to be responsible for generating the Higgs potential: in that case $g_* < g_{SM}$ would likely require a separate tuning anyways within the strong sector. Thus we will not consider the possibility of very small $g_*$ any further.  Instead we will have to be content to live with some amount of tuning (the specific implementation of the little hierarchy problem), which we quantify below. 
 
Clearly, the tuning here will be proportional to $v^2/f^2$. One simple way of quantifying it is to consider the magnitudes of the individual terms \footnote{
We write the Higgs potential as $V(h) = \sum_i \Delta V_i(h), \, \Delta V_i(h) = \frac{g_{SM,i}^2 \, \Lambda^2_i}{16 \pi^2} \left( -a_i |h|^2 + b_i \frac{|h|^4}{2f^2} \right)$.
}
that would contribute to a shift of the VEV of the Higgs 
\begin{equation}
\label{tuningv}
\Delta_{v^2} = \frac{\delta v^2}{v^2_{exp}} \sim \frac{1}{(246 \GeV)^2} \frac{a_i}{b_i} f^2 \ ,
\end{equation}
where $a_i$ and $b_i$ are the generic magnitudes of the terms appearing in the potential (which are then assumed to partially cancel against each other). Since this tuning involves 
the ratios of two terms generated in the potential (and since the magnitudes of the individual terms in the potential are known) it is better to instead separately consider the tuning in the mass term and the quartic term. The tuning for the mass parameter is
\begin{equation}
\label{tuningmu}
\Delta_{\mu^2} = \frac{\delta \mu^2}{\mu^2_{exp}} \sim \frac{1}{(800 \GeV)^2} a_i \, g_{SM,i}^2 \Lambda_i^2 \ ,
\end{equation}
while the tuning for the Higgs quartic is
\begin{equation}
\label{tuninglambda}
\Delta_{\lambda} = \frac{\delta \lambda}{\lambda_{exp}} \sim \frac{1}{3^2} b_i \, g_{SM,i}^2 g_{*,i}^2 \ ,
\end{equation}
where again $a_i$ and $b_i$ are the individual contributions to these terms before any cancellation. 
Notice that even in the most favorable situation for $\Delta_{\lambda}$, that is $g_{*} \simeq g_{SM}$, an irreducible tuning remains from the mass parameter, given that $\Delta_{\mu^2} \sim (f/270 \GeV)^2$, where we have taken $g_{SM}^2 \sim N_c y_t^2$, and experimentally $f > v$ is required.

An important consequence of this discussion is that since the Higgs mass determines the value of the Higgs quartic, it is no longer reasonable to assume an order one Higgs quartic (since we know if is fixed to $\lambda \approx 0.13$). One popular way of reducing the fine-tuning in composite Higgs models was to assume that while the mass parameter is generated at loop level, the quartic is generated at tree level (corresponding to $a\sim 1, b\sim (4\pi)^2$). This would eliminate the tuning in $v$ (due to the relation $v\sim f/ (4\pi )$), however now the quartic would come out too large, requiring in turn a tuning in $\lambda$ to reduce the Higgs mass to the observed value. 
 
We can summarize the discussion of the tuning in the Higgs potential in the following way: the experimental data suggests that both $\mu^2$ and $\lambda$ must be loop suppressed, and to minimize the tuning one would like $f$ being as close to $v$, and $g_{*}$ as close to $g_{SM}$, as possible.

\section{Classification of the Composite Higgs Models based on Higgs potential  \label{classification}}
\setcounter{equation}{0}
\setcounter{footnote}{0}

Based on the discussion of the previous section we can now classify the various types of composite Higgs models based on the  generic magnitudes of the Higgs mass and quartic parameters they would be predicting. 

\begin{center}
{\bf \large \noindent $\bullet$ Tree-level mass and quartic:  
$a = {\cal O} (1), b = {\cal O}(1),  g_{*} \sim 4\pi$  \\ ``Bona fide" composite Higgs}
\end{center}

These models can be regarded as Technicolor models with an enlarged global symmetry, the breaking of which yields an extra ``pion'' with the quantum numbers of the Higgs \cite{Galloway:2010bp}. However, they typically predict a too large Higgs mass term and quartic coupling, with generically $v \sim f$.
Even if $a$ is tuned by an amount $\sim \xi = v^2/f^2$, the Higgs is still too heavy, since 
$\lambda \sim g_{SM}^2 \sim N_c y_t^2$. Thus a second independent tuning must be made on $b$.
Overall, we can roughly estimate the tuning required in this class of models as \footnote{
We will be assuming for simplicity that two uncorrelated cancelations, one in $\mu^2$ and another in $\lambda$, take place.}
\begin{equation}
\Delta = \Delta_{\mu^2} \times \Delta_{\lambda} \sim (0.003 \, \%)^{-1} \left( \frac{f}{1 \TeV} \right)^2\ .
\end{equation}

\begin{center}
{\bf \large \noindent $\bullet$ Loop-level mass, tree-level quartic:  
$a = {\cal O} (1), b = {\cal O}(\frac{16\pi^2}{g_*^2}),  g_{*} \ll 4\pi$  \\ Little Higgs models}
\end{center}

The ``little'' Higgs models~\cite{ArkaniHamed:2002qx,ArkaniHamed:2002qy,Low:2002ws,Kaplan:2003uc,Chang:2003un,Skiba:2003yf,Schmaltz:2004de,Schmaltz:2010ac}  were invented to provide a fully natural Higgs potential: one automatically obtains a hierarchy between the Higgs VEV and $f$: $v^2/f^2 \simeq g_{*}^2/16\pi^2 \ll 1$, without tuning. This  however comes at the price of increasing the Higgs mass: since $\lambda \sim g_{SM}^2$, one would expect $m_h \sim 2 v g_{SM} \sim 500$ GeV for $g_{SM} \sim 1$. While a fully natural Higgs potential was very appealing before the value of the Higgs mass was known, once the Higgs mass is pinned down to 125 GeV one needs to perform additional tunings in $a$ and $b$ to obtain this mass.  Thus the Higgs potential in little Higgs theories can not be considered fully natural anymore. 
A naive estimate of the tuning involved is given by
\begin{equation}
\Delta = \Delta_{\mu^2} \times \Delta_{\lambda} \sim (1 \, \%)^{-1} \left( \frac{f}{1 \TeV} \right)^2 \left( \frac{g_*}{\sqrt{N_c} y_t} \right)^2\ .
\end{equation}
where we have taken $g_{SM} \sim 1$.\footnote{In most little Higgs models the leading quartic Higgs coupling is not generated from the same SM coupling than the mass term, the latter typically arising from top loops.}

The crucial ingredient that allows the Higgs mass parameter to become loop suppressed in little Higgs models is called collective symmetry breaking:  the Higgs doublet transforms under some extended global symmetry, which is not completely broken by any single interaction term. Since one needs a chain of these terms to feel the symmetry breaking, one loop diagrams will not be quadratically divergent, and hence will not be cut off by the naive scale of compositeness $\Lambda_C = 4\pi f$, but rather at some earlier scale. This lower scale is set by the masses of the light composite resonances, $m_{*} = g_{*} f$, which are called top partners for the top-loop, or vector partners for the gauge-loop. It is technically natural for the top/vector partners to be lighter than the strong coupling scale $\Lambda_C \sim 4 \pi f$, and in addition, the mechanism of partial compositeness, which we will discuss in detail in Section~\ref{yukawa}, naturally realizes light top partners for a sizable degree of compositeness of the top. Collective breaking then requires that  $g_{*}$ must not be much larger than $g_{SM}$. In particular, this fact implies that the top/vector partners must be weakly coupled. 

Little Higgs models are chosen such that the collective breaking protects the Higgs mass parameter hence $a={\cal O}(1)$, while a tree-level quartic is generated by means of extra scalars leading to $b={\cal O}(16\pi^2/g_*^2)$.\footnote{
These scalars get $\sim f/v$ larger mass terms than the Higgs, and can thus be consistently integrated out for what the Higgs potential concerns.}
The collective breaking mechanism also ensures that the large tree-level effective quartic does not lead to enhanced corrections to the Higgs mass term, the so-called collective quartic \cite{Schmaltz:2008vd}.

\begin{center}
{\bf \large \noindent $\bullet$ Loop-level mass and quartic:  
$a = {\cal O} (1), b = {\cal O}(1),  g_{*} \ll 4\pi$  \\ ``Holographic" composite Higgs}
\end{center}

This is the scenario where the entire Higgs potential is loop generated. These models need one tuning in the Higgs potential of order $\xi =v^2/f^2$ in order to achieve the right Higgs VEV $v< f$. However, once this tuning is achieved, the Higgs mass will automatically be light. Again the divergences in the Higgs potential are cut off at the scale of the top and vector partners. 
Thus, the generic tuning required in this case scales as
\begin{equation}
\Delta = \Delta_{\mu^2} \times \Delta_{\lambda} \sim (7 \, \%)^{-1} \left( \frac{f}{1 \TeV} \right)^2 \left( \frac{g_*}{\sqrt{N_c} y_t} \right)^4\ .
\end{equation}
where $g_{SM}^2 \sim N_c y_t^2$ has been taken.

These models were inspired by the AdS/CFT correspondence: some strongly interacting theories can be described by weakly coupled AdS duals.  The existence of such a dual  is intrinsically tied to the presence of ``weakly'' coupled resonances in the large $N$ regime, with coupling $g_{*} \sim 4 \pi / \sqrt{N}$. One can include in this class of models their deconstructed~\cite{ArkaniHamed:2001nc} versions as well, with several sites and links \cite{4dch}.

The holographic composite Higgs models also feature a version of collective breaking mechanism both in the gauge and fermion sectors, which is a consequence of extra-dimensional locality (or theory-space locality, its discrete version for the deconstructed case)~\cite{Thaler:2005kr}.
This protection is generically absent in the scalar sector for the holographic Higgs. However since the quartic is already loop-suppressed, the loop contribution to the Higgs mass from the Higgs self-interaction will be effectively two-loop suppressed, and hence is not dominating even if it is cut off at a scale higher than the top/vector partners.  The same will hold for contributions to the Higgs potential obtained from integrating out additional GB's.  Thus we can summarize the two main differences between little Higgs models and holographic composite Higgs models: little Higgs models feature a tree-level collective quartic $b = O(16\pi^2/g_{*}^2)$, generated from integrating out a particular class of ``heavy'' GB's \cite{Schmaltz:2008vd}, while holographic Higgs models have a loop suppressed quartic. Collective breaking in little Higgs models will ensure that the Higgs mass contribution from scalar and self-interactions is suppressed despite the appearance of a large effective quartic, while no such mechanism is at work in holographic models. In those models the quartic is simply small, thus also ensuring the appropriate suppression of the Higgs mass term. 

Then, the collective breaking in holographic Higgs models affects the Higgs mass term as well as the other pGB's, such that the Higgs is only lighter than these extra scalars at the expense of tuning the Higgs VEV, that is $m_h^2 \sim L v^2$ while $m_H^2 \sim L f^2$. This is in contrast with little Higgs models, where generically only the Higgs mass term is protected, but not the other pGB's, in particular those involved in the generation of the quartic Higgs coupling. The result in this case is $m_h^2 \sim g_{SM}^2 v^2$ while $m_H^2 \sim g_{SM}^2 f^2$, that is the same ratio than in holographic Higgs models but without the loop suppression.

\begin{center}
{\bf \large \noindent $\bullet$ Twin Higgs:  
$a = {\cal O} (1), b = {\cal O}(1)-{\cal O}(\frac{16\pi^2}{g_*^2}),  g_{*} = g_{SM}$ }
\end{center}

The ``twin'' Higgs models~\cite{twin} yield the same prediction for $a$ as little Higgs or holographic Higgs models, but the mechanism to eliminate the quadratic divergences in the Higgs mass term is based on a discrete $Z_2$ symmetry instead of collective breaking.
Regarding $b$, the generic prediction is a loop level Higgs quartic coupling, thus as in holographic Higgs models $b = {\cal O}(1)$, although when these models were originally proposed, it was convenient to introduce by hand a tree level quartic, such that $b = {\cal O}(16\pi^2/g_*^2)$ and a hierarchy $v < f$ was naturally generated, as in little Higgs models.
However, since the overall scale of the Higgs potential is now known, the latter option is no longer preferred, as discussed in Sec.~\ref{potential}.

The most important difference with respect to the previous models is that the partners cutting off the potential do not necessarily carry SM charges, in particular color. 
Given the lack of positive signals of top partners at the LHC, this is a relatively unexplored scenario in which opportunities for model building are still open, with the potential to produce interesting developments.

\begin{center}
{\bf \large \noindent $\bullet$ Dilatonic Higgs}
\end{center}

This scenario is quite different from the previous ones, and it is not very useful to compare them based on the form of the Higgs potential. In this case the dilaton (the pGB of spontaneously broken scale invariance) is playing the role of the 125 GeV Higgs-like particle~\cite{dilaton,Chacko:2012sy,Bellazzini:2012vz,naturallylight}. The analog state in the warped extra dimensional models is the radion~\cite{radion}, the studies of which have inspired much of the work in the general 4D framework. However, for these ``dilatonic'' Higgs models it is very important to point out that the dilaton VEV is not directly related to the electroweak VEV, or in other words $m_W^2 \neq g^2 \vev{h}^2/4$, unlike for a genuine Higgs.
Instead, the VEV of the dilaton actually fixes the overall scale of the potential, $\vev{h} \equiv f$, relative to a given UV scale $\mu_0$.
This explains why in the limit of exact scale invariance the dilaton potential only contains a quartic term (which itself is consistent with scale invariance).
A non-trivial minimum is then achieved due to explicit scale invariance breaking induced by the running couplings, which introduces an  implicit dependence of $g_{SM}$ on $h/\mu_0$, of the form $g_{SM} \sim (h/\mu_0)^{\gamma_{SM}}$,
where $\gamma_{SM}$ is the anomalous dimension associated to $g_{SM}$.  
Furthermore, a minimum with $f\ll\mu_0$ only arises naturally for $g_{SM} \sim 4 \pi$ at the condensation scale, which is commonly taken as an indication that the potential of the dilaton is driven by a non-SM coupling.\footnote{Although one possibility is that the coupling of the top to the strong sector, which is related to its Yukawa, drives the spontaneous breaking of scale invariance.}

In order for the dilaton to resemble the SM Higgs, $f$ must accidentally be close to $v$, for instance if only operators with the quantum numbers of the SM Higgs condense.
Therefore the experimental constraints in this case go in the opposite direction that in the previous models, pushing towards $v \sim f$.
Moreover, let us note that the dilaton could actually arise from a variety of scale invariant ``strong sectors", including those that are ``weakly'' coupled, that is $g_{*} \ll 4 \pi$.
However, explicit calculations using AdS/CFT imply that the large $N$ limit associated with this scenario is not preferred, since it tends to push $f \gg v$.\\

As a final remark in this section, we would like to empahsize that twisted versions of the models reviewed above also exist. For instance, due to constraints from electroweak precision constraints, which affect more significantly the boson sector of little Higgs models, it is known that it is favored not to extend the SM gauge group, at the expense of a collective symmetry breaking in the gauge sector that resembles that of holographic models.
This set-up was first proposed in \cite{Katz:2005au}, and later the littlest Higgs coset $\SU(5)/\SO(5)$ was realized a la holographic Higgs, first as a warped extra-dimensional model  in \cite{Thaler:2005en}, and then using the 4D effective description \cite{Vecchi:2013bja}.

As we have already done in this section, in the following we use the term ``partners'' to denote the new light and weakly coupled states that cut off the Higgs potential.

\section{Classification of the Composite Higgs Models based on Flavor Structure \label{yukawa}}
\setcounter{equation}{0}
\setcounter{footnote}{0}

Another important distinguishing feature of the various composite Higgs models is based on the mechanism for generating Yukawa couplings. The two main alternatives are condensation of 4-Fermi operators and partial compositeness. Further classification of the partially composite case can be done based on how the appropriate flavor hierarchies are actually achieved.

\subsection{Condensation of 4-Fermi operators}
\label{4Fermi}

This is the traditional way of obtaining Yukawa couplings in strongly coupled (Technicolor) theories~\cite{Dimopoulos:1979es}: a SM bilinear interacts with the strong sector,
\begin{equation}
\label{otc}
\lambda \bar \psi_L \psi_R \mathcal{O} \ ,
\end{equation}
where $\mathcal{O}$ is a scalar operator with the quantum numbers of the Higgs, for instance $\mathcal{O} = \bar \psi_{TC} \psi_{TC}$ in extended Technicolor models.
At low energies the operator $\mathcal{O}$ interpolates to a function of the Higgs, therefore giving rise to an ordinary Yukawa coupling of size
\begin{equation}
\label{ytc}
y_\psi \sim \lambda(\Lambda_F) \left(\frac{\Lambda_C}{\Lambda_F}\right)^{d-1} \ ,
\end{equation}
where $\lambda(\Lambda_F)$ is the value of the bilinear coupling at the flavor scale $\Lambda_F$, $\Lambda_C \sim 4 \pi f$ is the strong sector scale, and $d$ the dimension of the operator $\mathcal{O}$. This is the mechanism relied on in the bona-fide composite Higgs models. The most refined version of it goes under the name of conformal technicolor \cite{Luty:2004ye}, which tries to explain why the Higgs has properties similar to an elementary scalar in the Yukawa interactions where it is linearly coupled, but it is very different from an elementary scalar in the Higgs mass term where it appears quadratically. Conformal technicolor would assume that while the dimension of the linear Higgs operator is close to one, in order to allow a large enough $\Lambda_F$ as to satisfy flavor constraints while reproducing the sizable Yukawa of the top, that of the quadratic one is bigger than four, rendering it irrelevant. 
It also departs from the proposal of walking technicolor \cite{walking} in that the large-$N$ limit of the strong gauge group is not taken, to avoid large contributions to the $S$-parameter.  However the basic assumption is under stress from recent general bounds on scaling dimensions in 4D CFT's using conformal bootstrap \cite{bootstrap}.

\subsection{Partial compositeness}
\label{pc}

All the other composite Higgs models use the alternative mechanism for generating Yukawa couplings known as partial compositeness. Although this mechanism was originally proposed to address the flavor problem in Technicolor models~\cite{Kaplan:1991dc}, its power was not appreciated until its realization, via the AdS/CFT correspondence, as the localization of bulk fermions along a warped extra dimension in Randall-Sundrum models \cite{RSflavor,Agashe:2004cp,Agashe:2004rs}.
Here each SM fermion chirality couples to a different composite fermionic operator ${\cal O}_{L,R}$ of the strong sector,
\begin{equation}
\label{opc}
\lambda_L \bar{\psi}_L \mathcal{O}_R + \lambda_R \bar{\psi}_R \mathcal{O}_L \ .
\end{equation}
At low energies the state to be identified with the SM fermion is a mixture of $\psi_{L,R}$ and the lowest excitation of $\mathcal{O}_{L,R}$, which we call $\Psi_{L,R}$,
to be identified with the vectorlike fermionic partners of the SM fermions. The fraction of compositeness of the SM fields is characterized by the parameters $f_{L,R}$, which depend on the mixing matrices $\lambda_{L,R}$, as well as the fermionic composite spectrum, $m_{\Psi_{L,R}}$, as $f_{L,R} \simeq \lambda_{L,R} f/ m_{\Psi_{L,R}}$.  Assuming the Higgs is fully composite and has unsuppressed Yukawa couplings $Y_{u,d}$ with the composites $\Psi_{L,R}$, the effective SM Yukawa couplings $y_{u,d}$ for the SM fermions will be given by 
\begin{equation}
\label{ypc}
y_u^{ij} = f_{q}^i Y_u^{ij} f_u^{j} \, , \ \  y_d^{ij} = f_{q}^i Y_d^{ij} f_d^j \ .
\end{equation}
There are two main approaches to obtaining the correct flavor hierarchy without introducing large flavor violating interactions involving the SM fermions. If the composite sector has no flavor symmetry, then $Y_{u,d}$ are matrices with random ${\cal O}(1)$ elements. In this case a hierarchical structure in the mixing matrices $f_{L,R}$ can yield the right flavor hierarchies together with a strong flavor protection mechanism called RS-GIM. The other option is that the composite sector has a flavor symmetry, which would then be the source of the flavor protection. In this case some of the mixing matrices $f_{L,R}$ should be directly proportional to the SM Yukawas $y_{u,d}$.

\subsubsection{Anarchic Yukawa couplings}

The most popular version of partial compositeness is called the anarchic approach to flavor, where the underlying Yukawa couplings of the composites $Y_{u,d}$ are generic ${\cal O}(1)$ numbers without any structure. The flavor hierarchy in this case arises due to the hierarchical nature of the mixings between the elementary and the composite states $f_{L,R}$, due to large anomalous dimensions of the composite operators ${\cal O}_{L,R}$. In this case the mixing is expected to be given by 
\begin{equation}
f_{L,R}(\Lambda_C) \sim f_{L,R}(\Lambda_F) \left(\frac{\Lambda_C}{\Lambda_F}\right)^{d_{L,R}-5/2}
\end{equation}
where $d_{L,R}$ are the scaling dimensions of the composite operators, and $f_{L,R}(\Lambda_F)$ are the values of the mixing parameters at the flavor scale $\Lambda_F$. A hierarchical flavor structure arises naturally for ${\cal O}(1)$ anomalous dimensions.  The CKM mixing matrix arises from the diagonalization of the anarchic Yukawa matrices (\ref{ypc}) resulting in hierarchic left and right rotation matrices for the up and down sectors $L_u^{ij} \sim L_d^{ij} \sim {\rm min} (f_q^i/f_q^j, f_q^j/f_q^i), R_{u,d}^{ij} \sim {\rm min} (f_{u,d}^i/f_{u,d}^j, f_{u,d}^j/f_{u,d}^i)$. This results in a hierarchical  CKM matrix completely determined by the mixing of the LH states, and with relations $f_q^1/f_q^2 \sim \lambda, f_q^2/f_q^3 \sim \lambda^2, f_q^1/f_q^3 \sim \lambda^3$ (where $\lambda$ is the Cabibbo angle), while the diagonal quark masses are given by $m_{u,d}^i = f_q^i f_{u,d}^i v$. 

One of the consequences of this mechanism is that for states where the mixing is close to maximal, the mass of the heavy state must be well below the compositeness scale $\Lambda_C$. We can understand this by considering the interplay between a single composite fermion multiplet with mass $m_{\Psi} = g_{\Psi} f$ and its couplings $\lambda_{L,R}$ with the elementary  fermions $\psi_{L,R}$. The mixing parameter is given by 
\begin{equation}
f_{L,R} = \frac{\lambda_{L,R}}{\sqrt{\lambda_{L,R}^2 + g_{\Psi}^2}} \ .
\end{equation}
For this to approach unity we need $g_\Psi \ll 4\pi$, in agreement with our original expectation that the state responsible for cutting off the quadratic dependence of the Higgs potential should appear well below the cutoff scale. 

Flavor violations in this anarchic scenario are protected by the RS-GIM mechanism~\cite{Agashe:2004cp}, which is simply the fact that every flavor violation must go through the composite sector, thus all flavor violating operators will be suppressed by the appropriate mixing factors. For example, a typical $\Delta F=2$ 4-Fermi operator mediated by a composite resonance of mass $m_\rho$ and coupling $g_\rho$, will have the structure 
\begin{equation}
f_q^i f_q^{\dagger j} f_q^k f_q^{l \dagger} \frac{g_\rho^2}{m_\rho^2} \bar{q}^i q^j \bar{q}^k q^l \ ,
\label{fermiflavor}
\end{equation}
leading to a quark-mass dependent suppression of these operators. As we will review in Sec.~\ref{flavor}, the RS-GIM mechanism with completely anarchic Yukawa couplings is not sufficient to avoid the stringent flavor constraints from the Kaon system or from several dipole operators, pushing the compositeness scale $f$ to the multi-TeV regime.

\subsubsection{Flavor symmetries in the composite sector}

Another possible way of protecting the flavor sector from large corrections is by imposing a flavor symmetry on the composite sector. In this case we will lose the explanation of the origin of the flavor hierarchy, however might be able to obtain a setup that is minimally flavor violating (MFV), or next-to-minimally flavor violating (NMFV).   This was first carried out in the extra dimensional context in~\cite{Rattazzi:2000hs,CCGMTW,Santiago:2008vq}, and later implemented in the four dimensional language in~\cite{Redi:2011zi,Barbieri:2012tu}. The flavor symmetry structure is determined by the flavor structure of the mixing matrices $\lambda_{L,R}$  as well as the composite Yukawa matrices $Y_{u,d}$.  A flavor invariance of the composite sector will imply that the composite Yukawas are proportional to the unit matrix $Y_{u,d} \propto $ Id$_3$ for the case with maximal $\U(3)^3$ flavor symmetry in the composite sector. In order to have MFV, we need to make sure that the only sources of flavor violation are proportional to the SM Yukawa couplings. The simplest possibility is to make the LH mixing matrix proportional to the unit matrix, and the RH mixing matrices proportional to the up- and down-type SM Yukawa couplings
\begin{equation}
\lambda_L \propto {\rm Id}_3, \ \ \lambda_{Ru} \propto y_u, \ \ \lambda_{Rd} \propto y_d.
\label{LHcompositeness}
\end{equation}
This scenario corresponds to the case with composite left-handed quarks and elementary right-handed quarks, and an explicit implementation of MFV. However, the fact that the left-handed quarks are composite will imply potentially large corrections to electroweak precision observables. 
The other possibility is to introduce the flavor structure in the left handed mixing matrix. In order to be able to reproduce the full CKM structure, one needs to double the partners of the LH quarks to include $Q_u$ and $Q_d$: the composite Yukawa of $Q_u$ will give rise to up-type SM Yukawa couplings, while those of $Q_d$ to down-type Yukawas, while their mixings $\lambda_{Lu},\lambda_{Ld}$ are proportional to the SM Yukawas. Hence the ansatz for right-handed compositeness is 
\begin{equation}
\lambda_{Lu} \propto y_u, \ \ \lambda_{Ld} \propto y_d, \ \ \lambda_{Ru} \propto {\rm Id}_3, \ \ \lambda_{Rd} \propto {\rm Id}_3,
\label{RHcompositeness}
\end{equation}
which is also an implementation of MFV. \\

In the MFV scenarios discussed above the composite sector has a $\U(3)^3$ flavor symmetry, and either the LH or RH quarks are substantially composite, the degree fixed such as to reproduce the Yukawa coupling of the top. However, the light quarks appear to be very SM-like, more so after LHC dijet production measurements $pp \to jj$ in agreement with the SM, and it might be advantageous to reduce the flavor symmetry, allowing only the third generation quarks to be composites. Furthermore, the models with large flavor symmetries can significantly influence the predictions for the Higgs potential. If parts of the first and second generation are largely composite, along with that of the third, their contributions to the Higgs potential will be enhanced beyond the usual expectations.
Accordingly, the phenomenology of the fully MFV models can be significantly modified, as we comment in Sec.~\ref{signals}. A lot of effort has been put recently into exploring the models where the  third generation is split from the first two. This next-to-minimal flavor violation corresponds to imposing a $\U(2)^3 \times \U(1)^3$ or $\U(3)^2 \times \U(2) \times \U(1)$ flavor symmetry on the composite sector: it is phenomenologically viable or even favored \cite{Barbieri:2012uh,Barbieri:2012tu,Redi:2012uj}, keeping the natural expectations that the Higgs potential is saturated by the top and its partners. 
We will discuss the main phenomenological signatures of these scenarios in Sec.~\ref{flavor}.\\

Finally, there are other possibilities to reproduce the flavor structure of the SM while avoiding the constraints from flavor observables. These rely as well on flavor symmetries. One scenario, originally proposed in \cite{Rattazzi:2000hs}, is to assume that all the mixing matrices $\lambda_{L,R}$ are proportional to the identity, while all the flavor structure is provided by the composite sector, that is $Y_{u,d} \propto y_{u,d}$. This setup satisfies the rules of MFV, and all the SM quarks must have a large degree of compositeness.

One last logical possibility to comply with experiments is that the composite sector respects $CP$, given that most of the bounds come from $CP$-violating observables.
In this case the Yukawa couplings of the composite sector can be chosen to be real matrices, while the mixings introduce non-negligible $CP$ phases if the SM fermions are coupled to more than one composite operator. It has been shown in \cite{Redi:2011zi} that this idea might give rise to a realistic theory of flavor.
\\

\section{Cosets of symmetry breaking \label{cosets}}
\setcounter{equation}{0}
\setcounter{footnote}{0}

In this section we have compiled the most important symmetry breaking cosets $\Gsym/\Hsym$ from which a pseudo-Goldstone-Higgs could arise. The result is given in Table~\ref{tablecosets}.
Most of the global symmetry breaking patterns $\Gsym \to \Hsym$ have been described in the literature, mainly in the context of the little and holographic Higgs models.\\

\begin{table}[ht!]
    \begin{center}
\small{
	\begin{tabular}{cclccccc}
\hline

$\Gsym$ & $\Hsym$ & $C$ & $N_{G}$ & $\mathbf{r}_{\Hsym} = \mathbf{r}_{\SU(2) \times \SU(2)} \, (\mathbf{r}_{\SU(2) \times \U(1)})$ & Ref.\\

\hline

\SO(5) & \SO(4) & \checkmark & 4 & $\mathbf{4} = (\mathbf{2},\mathbf{2})$ & \cite{Agashe:2004rs} \\

$\SU(3) \times \U(1)$ & $\SU(2) \times \U(1)$ & & 5 & $\mathbf{2_{\pm 1/2}} + \mathbf{1_0}$ & \cite{Contino:2003ve,Schmaltz:2004de} \\

\SU(4) & \Sp(4) & \checkmark & 5 & $\mathbf{5} = (\mathbf{1},\mathbf{1}) + (\mathbf{2},\mathbf{2})$ & \cite{Katz:2005au,Gripaios:2009pe,Galloway:2010bp}\\

\SU(4) & $[\SU(2)]^2 \times \U(1)$ & $\checkmark^*$ & 8 & $\mathbf{(2,2)_{\pm2}} = 2 \cdot (\mathbf{2},\mathbf{2})$ & \cite{Mrazek:2011iu} \\

\SO(7) & \SO(6) & \checkmark & 6 & $\mathbf{6} = 2 \cdot (\mathbf{1},\mathbf{1}) + (\mathbf{2},\mathbf{2})$ & $-$ \\

\SO(7) & $\textrm{G}_2$ & $\checkmark^*$ & 7 & $\mathbf{7} = (\mathbf{1},\mathbf{3})+(\mathbf{2},\mathbf{2})$ & \cite{Chala:2012af} \\

\SO(7) & $\SO(5) \times \U(1)$ & $\checkmark^*$ & 10 & $\mathbf{10_0} = (\mathbf{3},\mathbf{1})+(\mathbf{1},\mathbf{3})+(\mathbf{2},\mathbf{2})$ & $-$\\

\SO(7) & $[\SU(2)]^3$ & $\checkmark^*$ & 12 & $(\mathbf{2},\mathbf{2},\mathbf{3}) = 3 \cdot (\mathbf{2},\mathbf{2})$ & $-$\\

\Sp(6) & $\Sp(4) \times \SU(2)$ & \checkmark & 8 & $(\mathbf{4},\mathbf{2}) = 2 \cdot (\mathbf{2},\mathbf{2})$ & \cite{Mrazek:2011iu}\\

\SU(5) & $\SU(4) \times \U(1)$ & $\checkmark^*$ & 8 & $\mathbf{4}_{-5} + \mathbf{\bar{4}_{+5}} = 2 \cdot (\mathbf{2},\mathbf{2})$  & \cite{Bertuzzo:2012ya}\\

\SU(5) & \SO(5) & $\checkmark^*$ & 14 & $\mathbf{14} = (\mathbf{3},\mathbf{3}) + (\mathbf{2},\mathbf{2}) + (\mathbf{1},\mathbf{1})$ & \cite{ArkaniHamed:2002qy,Katz:2005au,Vecchi:2013bja} \\

\SO(8) & \SO(7) & \checkmark & 7 & $\mathbf{7} = 3 \cdot (\mathbf{1},\mathbf{1}) + (\mathbf{2},\mathbf{2})$ & $-$ \\

\SO(9) & \SO(8) & \checkmark & 8 & $\mathbf{8} = 2 \cdot (\mathbf{2},\mathbf{2})$ & \cite{Bertuzzo:2012ya}\\

\SO(9) & $\SO(5) \times \SO(4)$ & $\checkmark^*$ & 20 & $(\mathbf{5},\mathbf{4}) = (\mathbf{2},\mathbf{2}) + (\mathbf{1}+\mathbf{3} , \mathbf{1}+\mathbf{3})$ & \cite{Chang:2003zn} \\

$[\SU(3)]^2$ & $\SU(3)$ &  & 8 & $\mathbf{8} = \mathbf{1_0} + \mathbf{2_{\pm 1/2}} + \mathbf{3_0}  $ & \cite{ArkaniHamed:2002qx} \\

$[\SO(5)]^2$ & $\SO(5)$ & $\checkmark^*$ & 10 & $\mathbf{10} = (\mathbf{1},\mathbf{3}) + (\mathbf{3},\mathbf{1}) + (\mathbf{2},\mathbf{2})$ & \cite{Chang:2003un} \\

$\SU(4) \times \U(1)$ & $\SU(3) \times \U(1)$ &  & 7 & $\mathbf{3_{-1/3}} + \mathbf{\bar{3}_{+1/3}} + \mathbf{1_{0}} = 3 \cdot \mathbf{1_0} + \mathbf{2_{\pm 1/2}}$ & \cite{Schmaltz:2004de,twin} \\

\SU(6) & \Sp(6) & $\checkmark^*$ & 14 & $\mathbf{14} = 2 \cdot (\mathbf{2},\mathbf{2}) + (\mathbf{1},\mathbf{3}) + 3 \cdot (\mathbf{1},\mathbf{1})$ & \cite{Low:2002ws,Katz:2005au} \\

$[\SO(6)]^2$ & \SO(6) & $\checkmark^*$ & 15 & $\mathbf{15} = (\mathbf{1},\mathbf{1}) + 2 \cdot (\mathbf{2},\mathbf{2}) + (\mathbf{3},\mathbf{1}) + (\mathbf{1},\mathbf{3})$ & \cite{Schmaltz:2010ac}\\

\hline
	\end{tabular} \\
	}
\caption{Symmetry breaking patterns $\Gsym \to \Hsym$ for Lie groups.  The third column denotes whether the breaking pattern incorporates custodial symmetry. The fourth column gives the dimension $N_G$ of the coset, while the fifth contains the representations of the GB's under $\Hsym$ and $\SO(4) \cong \SU(2)_L \times \SU(2)_R$ (or simply $\SU(2)_L \times \U(1)_Y$ if there is no custodial symmetry). In case of more than two $\SU(2)$'s in $\Hsym$ and several different possible decompositions we quote the one with largest number of bi-doublets.}
\label{tablecosets}
    \end{center}
\end{table} 

The minimal requirement on the global symmetries of the strong sector is that the unbroken $\Hsym$ must contain an $\SU(2) \times \U(1)$ subgroup, while the coset $\Gsym/\Hsym$ must contain a $\mathbf{2_{\pm1/2}}$ representation corresponding to the quantum numbers of the Higgs doublet under $\SU(2)_L \times \U(1)_Y$.
However, in order to protect the $T$-parameter from large corrections, one may instead require the unbroken $\Hsym$ to contain a larger ``custodial'' symmetry $\SO(4) \cong \SU(2)_L \times \SU(2)_R$ (which in turn contains the previous $\SU(2)_L \times \U(1)_Y$). This ensures that the actual custodial $\SU(2)_C$ is left unbroken after the Higgs gets its VEV, avoiding excessively large contributions to the $T$-parameter of order $\sim v^2/f^2$.
In this case  the coset must contain a 4-plet representation of $\SO(4)$ (that is a $\mathbf{4} = (\mathbf{2},\mathbf{2})$ of $\SU(2)_L \times \SU(2)_R$).
In Table~\ref{tablecosets} we have introduced the column $C$ to mark the cases with custodial symmetry $\Hsym \supset \SU(2) \times \SU(2)$,  with \checkmark, while for the cases with  only $\Hsym \supset \SU(2) \times \U(1)$ this column is left blank.
Notice however, that if there are GB's in addition to the single Higgs which are charged under $\SU(2) \times \SU(2)$, such as extra doublets or triplets (under either of the two $\SU(2)$'s), the  $\SU(2)_C$ does not generically remain unbroken when all the scalars get a VEV. In such a case $\SO(4)$ is not large enough, and extra $\SU(2)$'s or extra discrete symmetries are required to ensure an unbroken custodial symmetry. When there are additional $\SU(2)$'s, misaligned VEV's can be allowed if a large enough ``custodial'' symmetry is present for $\SU(2)_C$ to remain unbroken in the vacuum, while for the case with discrete symmetries, the extra parities must enforce vanishing VEV's for the additional scalars. We denote the cases without extra custodial protection with $\checkmark^*$.
Aside from symmetries, the effects of these additional GB's could instead be tamed by the introduction of additional gauge bosons that eat them. This would allow the suppression of the  dangerous violations of custodial symmetry if the corresponding gauge coupling can be taken large, effectively reducing the coset to a smaller one without the dangerous GB's (we also denote these cases with $\checkmark^*$).\\

Several additional comments are in order regarding Table~\ref{tablecosets}: 
\emph{i)} Beyond rank 3 this is an incomplete list for $\Gsym$'s. We do not intend to be exhaustive here.
\emph{ii)} Further cosets can be obtained stepwise from Table~\ref{tablecosets} via  $\Gsym \to \Hsym \to \Hsym' \to \cdots$.
\emph{iii)} ``Moose''-type models are obtained by combining several copies of the cosets in Table~\ref{tablecosets}. This is the case for instance of the minimal moose of~\cite{ArkaniHamed:2002qx}, given by $[\SU(3)^2/\SU(3)]^4$, and likewise for other mooses~\cite{Chang:2003un,Schmaltz:2004de}.
\emph{iv)} In little Higgs models it is customary to gauge a subgroup of $\Gsym$ beyond the SM $\SU(2)_L \times \U(1)_Y$, in order to implement the collective breaking in the gauge sector. Therefore, not all the GB's in Table~\ref{tablecosets} appear as physical states in the spectrum.
In this regard, the gauge collective breaking in holographic models becomes apparent by extending the symmetry structure, for instance from $\SO(5)/\SO(4)$ to $[\SO(5)]^2/\SO(5)$, and gauging a $\SO(4)$ subgroup on one of the factors (or sites), while the SM $\SU(2)_L \times \U(1)_Y$ is gauged on the other. We do not include these possibilities as separate entries in Table~\ref{tablecosets}.
\emph{v)} Finally, little Higgs models with $T$-parity \cite{Cheng:2003ju,Cheng:2004yc} typically require extra global symmetries (and its breaking) beyond the model without $T$-parity they are built from. For instance, the ``littlest'' Higgs model $\SU(5)/\SO(5)$ is extended with a $[\SU(2) \times \U(1)]^2/\SU(2) \times \U(1)$ in \cite{Pappadopulo:2010jx} (see \cite{Low:2004xc,Csaki:2008se} for other attempts). We do not include any of these extensions either in Table~\ref{tablecosets}.\\

It is understood that the global symmetries of the strong sector contain an unbroken $\SU(3)_C$ factor that is gauged by the SM strong interactions, that is $\Gsym \times \SU(3)_C$. However, several models have been proposed that include the color group in a non-trivial way \cite{Agashe:2004ci,Agashe:2005vg,Gripaios:2009dq,Frigerio:2011zg}. One of the main motivations of these models is to provide a rationale for the apparent unification of forces in the SM. By embedding $\SU(3)_C$ in a simple group along with $\SU(2)_L \times \U(1)_Y$ (for instance in $\SO(10)$, $\SU(4)_1 \times \SU(4)_2 \times P_{12}$, or $\SO(11)$), the central charges of the strong sector are the same for all the SM gauge interactions, thus ensuring that the differential running of the SM couplings remains the same than in the SM.\footnote{
Of course this feature could also be an accidental property of the strong sector in those cases where $\SU(3)_C$ is factored out.}
One of the main implications of these constructions is that some of the GB's carry color (aka leptoquarks or diquarks).\\

At this point, it is worth to note which of these symmetry breaking patterns could arise from fermion bilinear condensation $\vev{\psi \psi'}$ \cite{Witten:1983tx}.
The possible cosets are $[\SU(N)]^2/\SU(N)$, $\SU(N)/\SO(N)$, or $\SU(2N)/\Sp(2N)$, depending on the representation of $\psi, \psi'$ under the strong gauge group, complex, real, or pseudo-real, respectively.
This fact might be relevant when considering possible UV completions of the composite Higgs.\\

Let us end this section by noting that more exotic possibilities have also been considered for $\Gsym/\Hsym$, in particular non-compact Lie groups. Besides the case of the dilaton, corresponding to $\SO(4,2)/\textrm{ISO}(3,1)$, other possibilities such as $\SO(4,1)/\SO(4)$ have also been considered~\cite{rattazzitalk,Urbano:2013aoa}, although much less investigation has been devoted to these cases, mainly due to the expectation that their UV completion is non-unitary.

\subsection{The minimal model with custodial symmetry: $\SO(5)/\SO(4)$}

The $\SO(5)/\SO(4)$ is the minimal coset containing custodial $\SO(4) \cong \SU(2)_L \times \SU(2)_R$ symmetry that gives rise to a Higgs bi-doublet $(\mathbf{2},\mathbf{2})$. 
The $\SU(2)_L$ factor and the $\U(1)_Y$ inside $\SU(2)_R$ are gauged by the SM electroweak interactions.
Other models with larger cosets that also implement custodial symmetry reduce to  this one when the symmetry breaking interactions make the other GB's heavy (or they are gauged away).

\begin{figure}[t!]
\begin{center}
\includegraphics[width=3in]{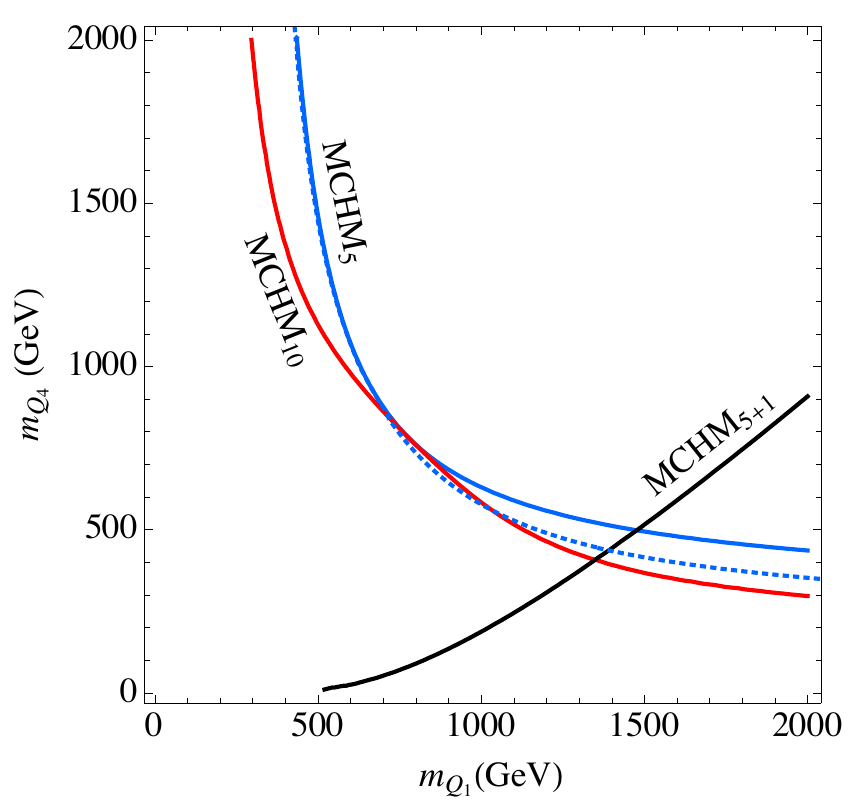}
\caption{Masses of the top partners $Q_1$ and $Q_4$ that reproduce the Higgs mass $m_h = 125 \GeV$ for $v^2/f^2 = 0.2$, from \cite{Pomarol:2012qf}. The different lines correspond to different $\SO(5)$ embeddings for the top quark. In blue $q_L, t_R \in \mathbf{5}$, in red $q_L, t_R \in \mathbf{10}$ (with $Q_1 \to Q_6$) and in black $q_L \in \mathbf{5}$ and $t_R \in \textbf{1}$.}
\label{PR}
\end{center}
\end{figure}

This model, whose origin can be traced back to~\cite{Chang:2003un} as a little Higgs moose model, and was realized as a warped extra-dimensional construction in~\cite{Agashe:2004rs} (MCHM), has been thoroughly examined in light of the Higgs discovery.
Besides the well-known fact that a certain degree of tuning is required to bring down $\mu^2$ to the observed value~\cite{tuningrefs,CFW} (see~\cite{Panico:2012uw} for a recent assessment), several approaches have been recently used to render the potential finite and therefore calculable, nailing down the features that the SM partners (top and electroweak) must have in order to reproduce the observations. Among these it is worth mentioning the ``moose'' extensions, either $\SO(5) \times \SO(5) / \SO(5)$ with extra $\SO(4)$ gauged \cite{Matsedonskyi:2012ym}, or $\SO(5) \times \SO(5) / \SO(5) \times \SO(4)$ with extra $\SO(5)$ gauged \cite{Redi:2012ha},
and the use of the Weinberg sum-rules (an old idea used to compute the pion masses in the QCD chiral Lagrangian)~\cite{Marzocca:2012zn,Pomarol:2012qf}.\footnote{It has also been shown in this set-up that extra colored vector resonances, or gluon partners, can mildly reduce the Higgs mass prediction via renormalization effects~\cite{Barnard:2013hka}.}
The conclusions of these works are similar to those previously obtained in realizations in a warped extra dimension~\cite{Contino:2006qr}, and which we have explained in Sec.~\ref{potential}: light and weakly coupled top partners are needed, and some tuning $\sim 5 \%$, is needed to push $f$ somewhat larger than $v$ and comply with the experimental constraints. We show in Fig.~\ref{PR} the plot from~\cite{Pomarol:2012qf} showing that at least one of the top partners (in a $\mathbf{1}$ and $\mathbf{4}$ representations of $\SO(4)$, with masses $m_{Q_1}$ and $m_{Q_4}$ respectively) must be light in order to reproduce the observed Higgs mass.\footnote{
Exceptions exist to this generic expectation~\cite{Marzocca:2012zn,Pomarol:2012qf}. These have been found in the context of a fully composite $t_R$, thus arising as a massless chiral composite. In this case $t_R$ does not contribute the Higgs potential, and the top Yukawa coupling is simply given by $y_t \simeq \lambda_L$, hence the degree of compositeness of $t_L$ is fixed.
Further, in these special cases the Higgs quartic is accidentally generated only at  $y_t^4$ order, instead of $y_t^2 g_{*}^2$, thus losing the connection small $\lambda$ $-$ small $g_{*}$. Hence the observed Higgs mass can be reproduced with heavier top partners. However, this is at the expense of increasing the tuning in $\mu^2$ (for fixed $f$), that scales as $y_t^2 g_{*}^2$, as expected.}\\

Let us conclude this section with another comment on the  $\sim 5 \%$ tuning in $\mu^2$. This tuning can be accomplished either by canceling two different top contributions, generically of ${\cal O}(\lambda_L^2)$ and ${\cal O}(\lambda_R^2)$, or by canceling the top versus the gauge contributions, of ${\cal O}(g^2)$. In this latter case the expectation is, confirmed in explicit constructions, that the top and gauge contributions appear with different signs, creating some degree of cancellation. Assuming that this is the case, the current upper bound on the gauge partner masses, $m_\rho \simeq 2.5 \TeV$ (see Sec.~\ref{ewpt}), gives us a direct clue on where the top partners should be: the approximate cancellation $N_c y_t^2 m_T^2 \simeq (9/8) g^2 m_\rho^2$ yields $m_T \simeq 1 \TeV$. This mass range will be thoroughly explored in the next phase of the LHC.

\section{Signals} \label{signals}
\setcounter{equation}{0}
\setcounter{footnote}{0}

The SM partners (new particles light compared to the cutoff $\Lambda_C \sim 4 \pi f$) play an important role in the generation of the Higgs potential in the little, holographic and twin Higgs scenarios, which can be considered the weakly coupled versions of the bona fide composite Higgs case.
The potential in these cases could be affected by large logs, $\log(\Lambda_C^2/m_{*}^2)$, where again $\Lambda_C$ is the compositeness scale while $m_{*}$ is a generic mass for the partners,
unless another layer of partners is light.
The partners, if present as suggested by the discussion in the previous section, generically give the leading contribution to electroweak precision tests (EWPT), in particular $S$, $T$, and $Z b \bar b$.
They can also give rise to important flavor transitions beyond the SM.
Also, they modify the couplings of the Higgs boson, to be taken into consideration along with the intrinsic deviations due to the composite nature of the Higgs.\footnote{Notice that the partners, being composite as it is the Higgs, will generically be affected by higher-dimensional operators, suppressed by suitable powers of $m_{*}/\Lambda_{C}$.}
Finally, such resonances should be produced at colliders, if sufficiently light and coupled to the SM matter.
All of these issues will be discussed in this section.\\

\subsection{Electroweak precision tests} \label{ewpt}

The electroweak precision observables characterize the properties of the SM gauge bosons, and their couplings to the SM fermions. Since we have not observed any particles beyond the standard model thus far, it is reasonable to assume that all new physics states are heavier than the electroweak scale.
This allows us, as a leading approximation, to parametrize their effects at the electroweak scale and below via higher dimensional operators with SM fields only.

\subsubsection{Universal}

Most of the new physics effects are of the ``universal type" and can be encoded in the modifications of the SM gauge bosons' two-point functions \cite{oblique}.
The most relevant effects in each class can be parametrized by the parameters \footnote{$\hat{S}$ and $\hat{T}$ are proportional to the Peskin-Takeuchi parameters $\hat{S}=g^2/(16\pi)S$ and $\hat{T}=\alpha_{EM}T$.} $\hat S$, $\hat T$, $W$, and $Y$, where the first two generically yield the most stringent constraints, since the other two are typically suppressed by extra powers of $g^2/g^2_*$. \\

There are two generic contributions to the $\hat{S}$ parameter which arise in all composite Higgs models: 
the UV contribution from heavy spin-1 resonances that can be estimated as
\begin{equation}
\hat{S}_{UV}\sim \frac{m_W^2}{m_\rho^2}\,,
\end{equation}
and an IR contribution associated with the reduced Higgs coupling $c_V$ to the EW gauge bosons \cite{Barbieri:2007bh}. This second one can be understood as follow. For $m_h\gg m_Z$, the $S$-parameter in the SM scales logarithmically with the Higgs mass as result of a cancellation of the log-divergent one-loop contributions of virtual Goldstone and Higgs bosons, $\log m_h/m_Z=\log\Lambda/m_Z -\log\Lambda/m_h$. 
 In composite Higgs models, while the Goldstone boson loop stays the same as in the SM, the Higgs boson loop is reduced and hence the cancellation is spoiled, leaving over $\log\Lambda/m_Z -c_V^2 \log\Lambda/m_h$. Thus the $S$-parameter becomes logarithmically sensitive to the new physics scale $\Lambda\sim m_\rho$ to be identified with the masses of the heavy resonances (of spin 0, 1, or 2) that couple to the $W$ and the $Z$ \cite{Barbieri:2007bh}
\begin{equation}
\label{SIR}
\hat{S}_{IR}\simeq \hat{S}_{SM}(m^{\text{eff}}_h)=\frac{g^2}{96\pi^2}\log\left(\frac{m_{h}^{\text{eff}}}{m_Z}\right)\,,\qquad m_{h}^{\text{eff}}=m_h\left(\frac{\Lambda}{m_h}\right)^{1-c_V^2}\,.
\end{equation}
Using a dispersion relation approach \cite{Orgogozo:2012ct} one can refine these estimates and achieve a $\mathcal{O}(m_h/m_\rho)$ accuracy in $\hat{S}$ at leading order in $g^2$ if the spectral density of the strong sector is known. For example, using vector meson dominance as in \cite{Marzocca:2012zn,Orgogozo:2012ct}, one finds
\begin{equation}
\hat{S}=\frac{g^2}{96\pi^2}\frac{v^2}{f^2}\left(\log\frac{m_\rho}{125\text{GeV}}-0.29\right)+ \frac{m_W^2}{f^2}\left(\frac{f_\rho^2}{m_\rho^2}-\frac{f_a^2}{m_a^2}\right) \ ,
\end{equation}
where $f_{\rho,a}$ and $m_{\rho,a}$ denote the decay constants and the masses of vector and axial resonances.
The new physics contribution to $\hat S$ can be kept under control if $m_\rho^2$ is sufficiently large, although this generically introduces some tuning in the Higgs potential, since $m_\rho^2$ fixes the scale where gauge-loop contributions are cut off.
Another option is to invoke some degree of cancellation between different contributions directly in $\hat S$, for instance coming from extra scalars of fermions \cite{Dugan:1991ck}, although these are loop suppressed and generically model dependent.\footnote{See e.g.~\cite{Barbieri:2008zt,Lodone:2008yy} for a discussion in the minimal $\SO(5)/\SO(4)$ model (MCHM).}
Moreover, in \cite{Matsedonskyi:2012ym} it was pointed out that fermion loops in composite Higgs models may provide additional sources of logarithmically enhanced contributions that can be understood in terms of the running of the two dimension-6 operators $\mathcal{O}_{W, B}$ related   to $\hat{S}$ \cite{Giudice:2007fh}. \\

It was recognized long ago \cite{Sikivie:1980hm} that the $\hat T$-parameter can be protected against new physics contributions by a custodial symmetry $\SU(2)_C \subset \SO(4) \cong\SU(2)_L \times \SU(2)_R$.
This requires that the new sector respects custodial symmetry to a very high degree, most often forbidding new sources of breaking beyond those already present in the SM, that is the Yukawa coupling of the top and the hypercharge gauge coupling.
In particular, it is required that the new states cutting off the Higgs potential, in particular the vector partners, come in complete representations of $\SO(4)$.
This has been explicitly verified in many little Higgs models, see for instance \cite{lhewpt}.
In holographic Higgs models this requirement is satisfied by construction, since the partners always come in complete representations of the unbroken global symmetry subgroup, which contains $\SO(4)$ \cite{holoewpt}.
In addition, while the custodial $\SO(4)$ is sufficient to protect the $\hat T$-parameter when a single Higgs field breaks the electroweak symmetry spontaneously, as we discussed in Sec.~\ref{cosets} this is not the case when extra scalar fields charged under $\SO(4)$ are present,  additional Higgs doublets, triplets, etc.
In these cases, an ``enlarged'' custodial symmetry is required (see \cite{Mrazek:2011iu} for a detailed explanation of the THDM case).

\begin{figure}[thb]
\begin{center}
\includegraphics[width=3in]{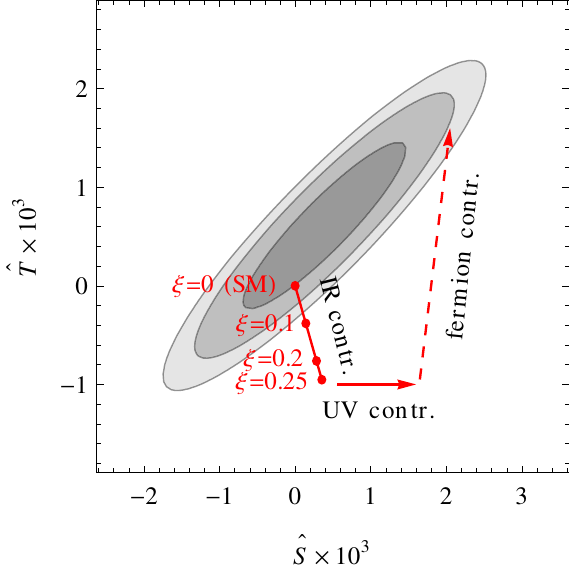}
\caption{Confidence level contours (at $65\%$, $95\%$ and $99\%$) for $\hat{S}$ and $\hat{T}$ from \cite{Grojean:2013qca}. The IR contributions alone would imply $\xi=v^2/f^2\lesssim 0.1$.}
\label{STplane}
\end{center}
\end{figure}

With custodial protection, the leading corrections to $\hat{T}$ arise thus at one loop. Analogously to the case for $\hat S$, there is a universal IR contribution from the reduced coupling of the Higgs boson which can again be estimated in the heavy Higgs limit as
\begin{equation}
\label{TIR}
\hat{T}_{IR}\simeq -\frac{3g^{\prime\,2}}{32\pi^2}\log\left[\frac{m_h}{m_Z}\left(\frac{\Lambda}{m_h}\right)^{1-c_V^2}\right] \ .
\end{equation}  
These IR contributions due to the modified Higgs couplings, eq.~(\ref{SIR}) and (\ref{TIR}), form a line in the $\hat{S}-\hat{T}$ plane. If these were the only corrections, then they would imply $\xi=v^2/f^2\lesssim 0.1$, see Fig.~\ref{STplane} reproduced from \cite{Grojean:2013qca}.

The one-loop contribution from fermions can be even more important: within the framework of partial compositeness it is generated by insertions of the mixings $\lambda_{L,R}$ and estimated as \cite{Giudice:2007fh}
\begin{equation}
\hat{T}_{fermions}\sim \frac{N_c}{16\pi^2}\frac{\lambda_L^4 f^2}{m_\Psi^2}\frac{v^2}{f^2}\,.
\end{equation}
which can be the leading contribution. See e.g. \cite{Matsedonskyi:2012ym,Barbieri:2008zt,Lodone:2008yy} about concrete realizations and examples. 
The above expression corresponds to the leading term in an expansion in $\lambda_{L}/g_{\Psi}$. However, if the degree of compositeness of the LH or RH top quark is large, the contributions to $\hat T$ are actually controlled by $m_\Psi$ \cite{Giudice:2007fh}. In that case $\hat T$ scales as $m_\Psi^2/m_\rho^2$, and it has been shown that such contributions can be positive for moderate values of $m_\Psi \sim 1 \TeV$ \cite{Pomarol:2008bh}. 

As shown in Fig.~\ref{STplane}, these contributions to $\hat T$ can be very important in order to bring the model into the $\hat S - \hat T$ ellipse and thus reduce the bound on $f$.

\subsubsection{Non-universal}

Besides the oblique parameters, strongly interacting models usually induce non-universal modifications to the couplings of the top, and due to $\SU(2)_L$ invariance, also to those of the left-handed bottom \cite{zbb}.
This is due to the necessarily large coupling of the top quark to the strong sector, in order to reproduce its large Yukawa coupling.
The strongest constraints come from measurements of the $Z b_L \bar b_L$ coupling, sensitive to the masses of the new-physics states.
However, it was shown in \cite{Agashe:2006at} that the $Z b_L \bar b_L$ vertex can be protected from large corrections by a $P_{LR}$ parity symmetry, as long as the $b_L$ embedding does not break it, that is if $b_L$ has $-1/2$ charge under both $\SU(2)_L$ and $\SU(2)_R$.\footnote{Notice that in symmetry breaking cosets with unbroken $\SO(4)$, $P_{LR}$ actually arises as an accidental symmetry of the leading order derivative Lagrangian \cite{Mrazek:2011iu}.}
As for the custodial symmetry, when this custodial parity is preserved by the strong sector, corrections to $Z b_L \bar b_L$ can be kept under control. 
Both symmetries yield important consequences for the quantum numbers and spectrum of the top partner resonances (for instance extended representations such as the $\mathbf{4} = (\mathbf{2},\mathbf{2})$).

\begin{figure}[thb]
\begin{center}
\includegraphics[width=3in]{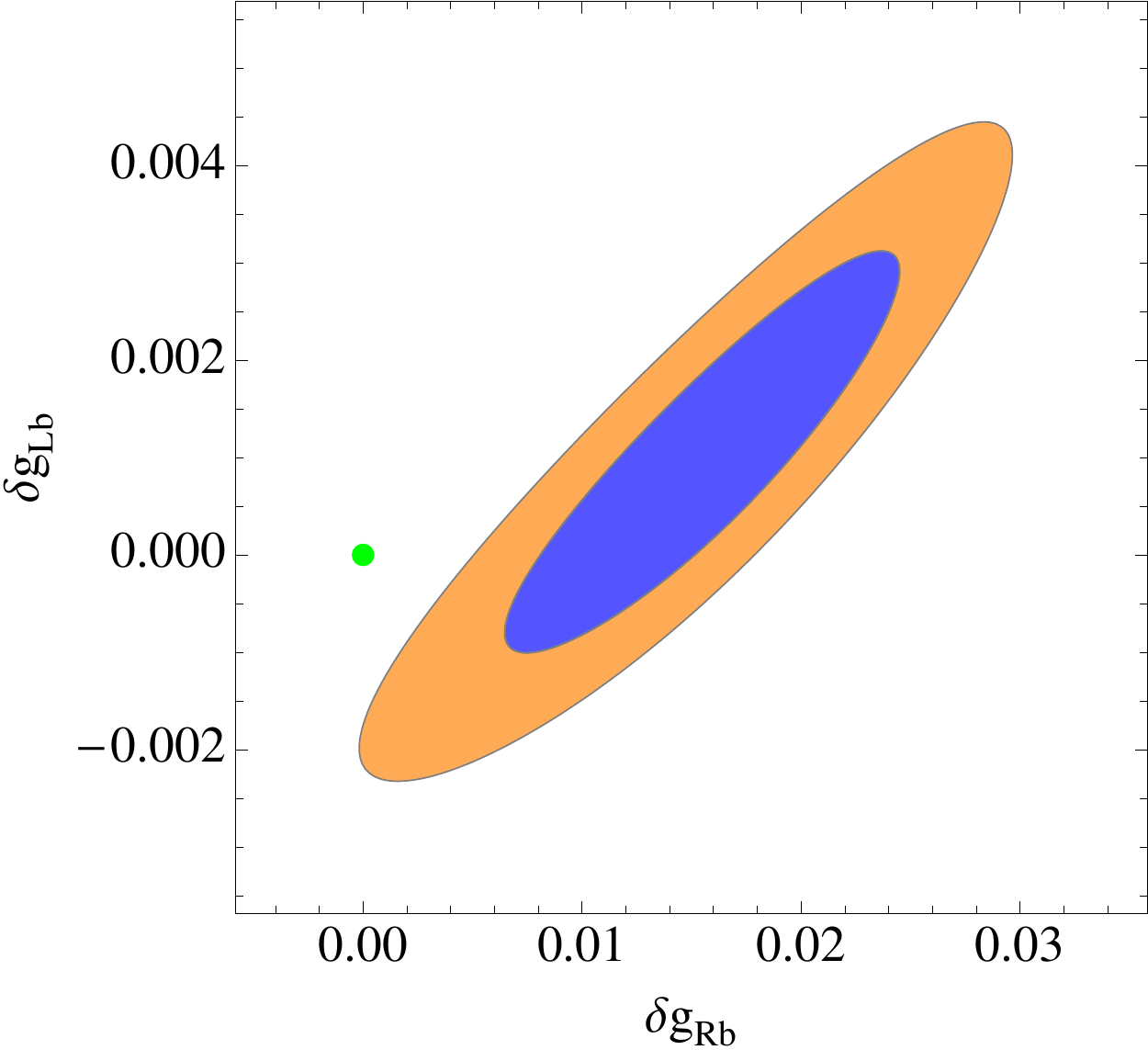}
\caption{Best fit region for the $Z\bar{b}b$ couplings from \cite{Batell:2012ca} favoring small positive $\delta g_{Rb}$. The SM is represented by the green point.}
\label{Zbbfigure}
\end{center}
\end{figure}

Fig.~(\ref{Zbbfigure}) reproduced from \cite{Batell:2012ca} shows the best fit region with a small positive $\delta g_{Rb}$ where the following parametrization is used \footnote{There exists another best fit region with a larger negative $\delta g_{Rb}$.}
\begin{equation}
\mathcal{L}=\frac{g}{c_W} Z_\mu \bar{b}\gamma^\mu\left[(g^{SM}_{Lb}+\delta g_{LB})P_L+(g^{SM}_{Rb}+\delta g_{RB})P_R\right]b\,.
\end{equation}

The contribution from fermion loops to $\delta g_{Lb}$ is generically logarithmically divergent as a result of insertions of the mixings that break the $P_{LR}$ parity
\begin{equation}
\frac{\delta g_{Lb}}{g^{SM}_{Lb}}\sim \frac{y_t^2}{16\pi^2}\frac{v^2}{f^2}\log\frac{\Lambda^2}{m_{\Psi}^2}\,.
\end{equation}

Another sensitive test concerns the anomalous coupling of the right-handed top and bottom to the $Z$.
This coupling is tightly constrained by $b \to s \gamma$ measurements. 
However, the size of the anomalous coupling is generically suppressed by $y_b/y_t$, yielding mild bounds on the new physics scale, see for instance \cite{Vignaroli:2012si}.
Other top related measurements still lack of precision \cite{Pomarol:2008bh,AguilarSaavedra:2011ct}.

In the previous sections we have argued that due to its contribution to the Higgs potential, fermionic top partners should be the lightest new physics states.
The effects of top-partners on precision tests, which we have reviewed in this section, have been thoroughly discussed in the literature, either in the context of little Higgs models \cite{Berger:2012ec}, holographic Higgs models \cite{Barbieri:2007bh,Pomarol:2008bh,Gillioz:2008hs,Lodone:2008yy,Anastasiou:2009rv,Barbieri:2012tu}, or in more generality \cite{Cacciapaglia:2010vn,Grojean:2013qca}.\\

Finally, let us again note that modified Higgs couplings to electroweak gauge bosons can be indirectly probed through electroweak precision measurements, \eq{SIR} and \eq{TIR}. Such modified couplings arise whenever the operator $(\partial_\mu (H^\dagger H))^2$ is generated, to which new physics contributes even if the states responsible for taming the Higgs potential only couple to the Higgs (even if they do not carry electroweak charges in particular).
Besides, this operator generically encodes the non-linear self-interations of the Higgs, intrinsic of its GB nature. As such, it will be suppressed by $\alpha/f^{2}$, with $\alpha$ a numerical factor that depends on the coset structure.

Also note that the case of a dilatonic Higgs needs to be considered separately for the EWPT's. Since a composite Higgs-like dilaton is not embedded into a SU(2) doublet, the argument before doed not directly apply.  Actually, the couplings of the dilaton to the gauge fields agree with those of the SM Higgs, except for a $v/f$ suppression. Thus the corrections to $\hat S_{IR}$ and $\hat T_{IR}$ are minimized in the limit $v/f \to 1$, the opposite limit than in ordinary composite Higgs scenarios.\\

For a recent model independent analysis of the constraints from EWPT, see \cite{Ciuchini:2013pca}.\\

Another important direction for taming electroweak precision constraints has been the introduction of T-parity~\cite{Cheng:2003ju,Cheng:2004yc}: a Z$_2$ discrete symmetry under which all BSM states are odd. Such a symmetry ensures that all corrections to electroweak precision observables from the new states are at least one loop suppressed, thus reducing the bounds on the masses of the new states. In this case one can obtain a theory consistent with the electroweak precision observables, even with new states as light as $\sim 1 \TeV$. T-parity has been one of the leading themes for Little Higgs models, and can of course also be implemented in the general 4D versions.\footnote{In warped extra-dimensional models one can find constructions with KK parity \cite{Agashe:2007jb}, which also aim at reducing the tension with electroweak precision measurement.} An illustration of the electroweak precision observables in a little Higgs model with T-parity can be found in~\cite{MaximJayPatrick}.

\subsection{Flavor and $CP$ violation} \label{flavor}

The interplay between electroweak symmetry breaking and the generation of the SM flavor structures has always been one of the major concerns in composite Higgs models.
The degree of the problem, and thus the importance of the constraints, can be understood by the number and expected size of the flavor structures present in the SM low-energy effective theory. This crucially depends on the mechanism employed to generate the SM Yukawas (see Sec.~\ref{yukawa}).

\subsubsection{4-Fermi operators}

It has been long known that a simple mechanism to generate the interactions in \eq{otc} gives rise also to unsuppressed SM flavor violating 4-Fermi interactions
\begin{equation}
\frac{c^{ijkl}}{\Lambda_F^2} {q}_i q_j \bar{q}_k \bar{q}_l
\label{flavor4Fermi}
\end{equation}
which generically violate the stringent flavor constraints: for instance from the Kaon system, $\Lambda_F>10^{3-5} \TeV$, while allowing for a sufficiently large top mass one would need  $\Lambda_F = {\cal O}(10) \TeV$.
As explained in Sec.~\ref{4Fermi}, this tension can be relaxed if the dimension of the operator ${\cal O}$ in \eq{otc} is sufficiently close to one, as long as the dimension of ${\cal O}^2$ does not decrease below four hence reintroducing the hierarchy problem.

It is worth mentioning that other alternatives might be viable, which rely on the flavor dynamics inducing additional suppression of the operators in \eq{flavor4Fermi}, either via the Yukawa couplings, $c^{ijkl} \sim y^{ij}_{u,d} \, y^{kl}_{u,d}$, in which case the bounds on $\Lambda_F$ can be relaxed close to the scale required to reproduce the top mass, or effectively imposing MFV, which could be realized if the couplings of the standard model fermions to the strong dynamics arise from the exchange of (supersymmetric) heavy scalars, such as in bosonic technicolor \cite{bosonictc}.
In the former case new physics is to be expected in flavor transitions, while in the latter supersymmetric states remnant of the flavor generation should be observable.

\subsubsection{Anarchic partial compositeness}

As discussed in Sec.~\ref{pc}, the RS-GIM mechanism of partial compositeness significantly reduces the contributions to dangerous flavor transitions. However, it has been shown that the suppression is not quite enough as to provide a fully realistic theory of flavor. Even though $\Delta F = 2$ 4-Fermi operators 
\begin{equation}
f_q^i f_q^{\dagger j} f_q^k f_q^{l \dagger} \frac{g_\rho^2}{m_\rho^2} \bar{q}^i q^j \bar{q}^k q^l
\label{fermiflavor2}
\end{equation}
are effectively suppressed by four powers of the fermion masses $m/v$ or CKM entries $V_{CKM}$, measurements of $CP$ violation in the Kaon system, $\epsilon_K$, put stringent bounds on the LR operators in \eq{fermiflavor2}, of the form $m_\rho \gtrsim 10 \frac{g_\rho}{Y_d}$ TeV~\cite{Agashe:2004cp,CFW,BurasMatthias,Redi:2011zi,KerenZur:2012fr}, as well on LL operators. Although less significant, qualitatively similar bounds on LL operators arise from $CP$ violation in the $B$ system, $m_\rho \gtrsim 1 \frac{g_\rho}{Y_u} \TeV$.
Given the expectation $m_\rho \sim g_\rho f$, these type of constraints bound the combination $Y_{d,u} f$. In explicit constructions of the pGB Higgs, the composite Yukawas $Y_{u,d}$ are correlated with the masses of the composite fermions cutting off the Higgs potential.
These kind of bounds therefore have a significant impact on the fine-tuning.
In addition, these bounds have to be contrasted with other potentially problematic flavor observables such as dipole operators 
\begin{equation}
f_q^i f_q^{\dagger j} \frac{1}{16 \pi^2} \frac{Y^3_{u,d}}{m_\Psi^2} \bar q_{i} \sigma_{\mu \nu} F^{\mu \nu} q_{j}
\end{equation}
generated by loops of composite fermions of mass $m_\Psi$ and the Higgs. These induce large contributions to $b\to s\gamma$, direct $CP$ violation in $\epsilon'/\epsilon_K$, and to the flavor conserving electric dipole moment of the neutron, all of them scaling with positive powers of $Y_d$, thus the constraints $m_\Psi > \alpha Y_d \TeV$, with $\alpha \sim 0.5 - 2$ \cite{Gedalia:2009ws,Altmannshofer:2011gn,Redi:2011zi,Barbieri:2012tu,KerenZur:2012fr}. All these flavor bounds taken together force the scale of compositeness to be above $~2 \TeV$ along with composite couplings $g_{*=\rho,\Psi} \gg g_{SM}$.\footnote{Other relevant effects which could also give rise to important constraints on $m_\Psi$ arise from flavor transitions mediated by the $Z$ \cite{Buras:2011ph}.}\\

Moreover, let us notice that the operators in \eq{fermiflavor2} could also be mediated by the Higgs or other pGB's (of mass $m_H$), with the associated enhancement of their coefficients by $(m_\rho^2/m_h^2)(v/f)^4$ or $(m_\rho^2/m_H^2)$, respectively.\footnote{Higgs mediated FCNC's will arise from the operators $\bar q_{i} H q_{j} H^\dagger H$.}
However it was pointed out in \cite{Agashe:2009di,Mrazek:2011iu} that these unwanted effects can be avoided thanks to the Goldstone nature of these scalars, as long as the embedding of the SM fermions into the global symmetries of the strong sector only allows for a single Yukawa-type operator $\bar q_L q_R F(h,H)$, thus enforcing MFV structure in the scalar interactions.\footnote{Flavor transitions mediated by extra pGB's can also be suppressed by forbidding their couplings to fermions via symmetries \cite{Glashow:1976nt,Buras:2010mh}.}

This is however not the case for a dilatonic Higgs, since the lack of a direct connection with the electroweak VEV generically implies that the fermion mass matrices and the dilaton couplings are misaligned. In that case the best alternative is to assume that the composite sector is endowed with flavor symmetries.\\

Let us briefly comment on the lepton sector. First of all, given that neutrinos are much lighter than charged leptons, and their mixings are not hierarchical, it is certainly plausible that neutrino masses come from a different source, or enjoy a different generation mechanism.
Factoring out the discussion of neutrino mass generation, the constraints on partial compositeness for leptons with anarchic Yukawas come from \cite{Agashe:2006iy,Csaki:2010aj,KerenZur:2012fr} 
the electron EDM, and $\mu \to e \gamma$ transitions from penguin mediated dipole operators. The bounds from experimental data are even more stringent than in the quark sector, which makes the minimal implementation of leptonic partial compositeness not viable. 

The most appealing option thus seems to rely on lepton flavor global symmetries, enforcing LMFV \cite{Redi:2013pga}. 
Another option to remove tree-level constraints on lepton partial compositeness is by imposing an A4 symmetry on the composite sector \cite{Csaki:2008qq}, alleviating the tension with the loop induced processes.
In that case the degree of compositeness of the leptons must increase in order to yield the proper Yukawa couplings, with the consequence of light tau partners \cite{delAguila:2010vg}.

\subsubsection{$\U(3)^3$ symmetric partial compositeness}

As review in Sec.~\ref{pc}, the scenarios falling into this category can be classified as LH or RH quark compositeness. The degree of compositeness in each case is fixed by the requirement $f_{L,R} \gtrsim y_t/Y_u$, in order to reproduce the top mass.
Therefore, in every case the inevitable signal will come from flavor diagonal 4-quark operators,
\begin{equation}
\frac{g_\rho^2}{m_\rho^2} f_{L,R}^4 (\bar{q} \gamma_\mu q) (\bar{q}' \gamma^\mu q')
\end{equation}
generated from the exchange of heavy resonances of mass $m_\rho$ and coupling $g_\rho$.
These have been recently probed at the LHC in $pp \to jj$ angular distributions. The individual bounds for the complete set of independent 4-quark operators, their coefficient normalized to $\Lambda^{-2}$, range between $\Lambda \gtrsim 1 -5 \TeV$ \cite{Domenech:2012ai}. These place strong constraints on the degree of compositeness of the quarks, given the identification $\Lambda \sim f/f_{L,R}^2$, for $m_\rho \sim g_\rho f$. Taking the most favorable situation, that is $f_{L,R} \sim y_t/Y_u$, the dijets constraints bound the combination $Y_u^2 f$, again implying large partners masses as in the anarchic case.

There is another class of constraints that apply only to LH or RH compositeness. If the LH quarks are composite, their (flavor diagonal) couplings to $W$ and $Z$ receive significant corrections, which affect precision observables such as quark-lepton universality in Kaon and $\beta$-decays or the hadronic width of the $Z$ \cite{Redi:2011zi}.\footnote{If the compositeness fraction of the LH leptons is equal to that of the LH quarks, there will be universal shifts in couplings to gauge bosons, which can be interpreted as a (too large) contribution to the $S$-parameter.} The corresponding bounds take the form $m_\Psi \gtrsim 35 f_L Y_u v$, which again, taking $f_{L,R} \sim y_t/Y_u$, implies a strong bound on the partners masses $m_\Psi \gtrsim 35 m_t$.
For the case of RH composite quarks, given that their coupling to $W$ and $Z$ are still poorly measured (and can be easily protected by their proper embedding into the global symmetries of the strong sector), the previous measurements do not yield important constraints. However, flavor violating LL 4-Fermi operators \eq{fermiflavor2} are still generated with a significant coefficient $~ (y_u y_u^\dagger)^2/(f^2 Y_u^4 f_R^4)$ \cite{Barbieri:2012tu}, which even though MFV suppressed, still yields $Y_u^2 f_R^2 f \gtrsim 6 \TeV$. Notice in particular that while this constraint prefers $f_R$ large, the dijet bounds push towards $f_R$ small.

In summary, flavor models with $\U(3)^3$ symmetry are under a significant stress from recent measurements of dijet production at the LHC. With the increase of energy at the next run of the LHC, such measurements will provide conclusive results about this possibility.

\subsubsection{$\U(2)^3$ symmetric partial compositeness and variants}

In models where the flavor symmetry is reduced in order to uncouple the fraction of compositeness of the light generations and that of the top quark, the compositeness constraints from measurements of $W$ and $Z$ couplings or dijet production (discussed above), are irrelevant. 
Therefore in these scenarios the only phenomenologically relevant flavor constraints are the consequences of the third generation (LH chirality, RH, or both) being distinct from the first two.
In this case it is important to point out that the R rotation matrices are very close to the identity in all the scenarios, with the corresponding suppression of the most dangerous LR 4-Fermi operators in \eq{fermiflavor2} \cite{Barbieri:2012tu,Redi:2012uj}.
Still the most sensitive flavor observables come from the Kaon and $B$ systems (and the $D$ system in the case of RH compositeness), as in the anarchic case, but with correlations among them, depending on the particular symmetry implementation. Most importantly, the associated bounds can now be satisfied for relatively low values of $f$ or the partner masses. This makes the $\U(2)$ scenarios the most favored ones for a natural electroweak scale, while still offering good prospects of new physics effects in flavor physics.\\

Let us conclude this section by commenting on the particulars of little Higgs models. Although their UV completion is not a priori determined, thus making an assessment of flavor and $CP$ violation more model dependent, solely from the interactions of the low energy degrees of freedom valuable lessons can be inferred, which are of course similar to those discussed in this section. Gauge and top partners contribute to neutral meson mixing and $CP$ violation, with bounds at the same level or in some cases milder than those coming from EWPT \cite{lhflavor,lhtparityflavor} (and see \cite{Berger:2012ec} for a recent review on the top partners effects).

\subsection{Higgs production and decay} \label{higgscouplings}

Higgs physics is a direct probe of the electroweak symmetry breaking sector, making the measurement and study of its couplings one of the major goals in particle physics today. This is particularly relevant in the composite Higgs scenario, given that its GB nature unavoidably implies non-linearities in its couplings to SM fields, i.e.~corrections of order $v^2/f^2$ with respect to the SM predictions. Importantly, this is regardless of any new states that might be present in the spectrum, given that such GB effects can not be decoupled.

\subsubsection{Single-Higgs production}

After the Higgs discovery, one of the major enterprises in particle physics has been the extraction of the linear couplings of the Higgs to the other SM fields.
These are obtained by fitting the experimental data on $\sigma \times BR$, see \cite{Azatov:2012qz,Falkowski:2013dza,Giardino:2013bma} and references therein.
The best tested Higgs couplings to date are those to electroweak gauge bosons $hZZ$ and $hWW$ (with less precision), and to massless gauge bosons $hgg$ and $h\gamma \gamma$, induced at one loop in the SM. Indirectly, through its contribution to $hgg$ and $h\gamma \gamma$, the coupling to top quarks, $h t \bar t$ is also being tested.
The first results on the coupling to tau leptons $h \tau \bar \tau$ and bottom quarks $h b \bar b$ have also been obtained. 

In order to make connection with the experimental data and compare with different models, we parametrize the linear interactions of the Higgs by the following Lagrangian
\begin{eqnarray}
\mathcal{L}_{eff}^{(h)} &=&
\left( c_V \left( 2 m_W^2 W^{+}_\mu W^{-\mu} + m_Z^2 Z_\mu^2 \right) - c_t m_{t} \bar{t} t  - c_b m_{b} \bar{b} b - c_\tau m_{\tau} \bar{\tau} \tau \right) \frac{h}{v} \nn \\
&& + \left( \frac{c_{\gamma\gamma}}{2} A_{\mu \nu} A^{\mu \nu} + c_{Z\gamma} Z_{\mu \nu} \gamma^{\mu \nu} + \frac{c_{gg} }{2} G^{a}_{\mu \nu} G^{a,\mu \nu}\right) \frac{h}{v}
\, ,
\label{Leff}
\end{eqnarray}
and present in Table~\ref{hlinear} the predictions for two distinct composite Higgs models, the $\SO(5)/\SO(4)$ model of \cite{Agashe:2004rs}, known as the Minimal Composite Higgs Model (MCHM), and the dilatonic Higgs following \cite{Bellazzini:2012vz}. For the MCHM, we only include the predictions associated to the GB non-linear nature of the Higgs, dictated by the symmetry structure of the model, and comment on the effects of the light SM partners below, which in any case give subleading corrections. For the case of the dilaton the couplings are entirely determined by scale invariance and its breaking.
\begin{table}[t!]
\centering
\begin{tabular}{cccc}
\hline
coupling & SM & MCHM & Dilaton \\ \hline
$c_V$ & 1 & $\sqrt{1-\xi}$ & $\sqrt \xi$ \\ 
$c_{\psi}$ & 1 & $\frac{1-(1+n_\psi) \xi}{\sqrt{1-\xi}}$ & $(1+\gamma_\psi) \sqrt \xi$ \\ 
$c_{\gamma \gamma}$ & 0 & 0 & $\frac{\alpha}{4 \pi} ( b_{IR}^{(EM)}-b_{UV}^{(EM)} ) \sqrt \xi$ 
\vspace{0.1cm} \\ 
$c_{Z \gamma}$ & 0 & 0 & 
$\frac{\alpha}{4 \pi t_W} ( b_{IR}^{(2)}-b_{UV}^{(2)} ) \sqrt \xi$
\vspace{0.1cm} \\ 
$c_{gg}$ & 0 & 0 & $\frac{\alpha_s}{4 \pi} ( b_{IR}^{(3)}-b_{UV}^{(3)} ) \sqrt \xi$
\vspace{0.05cm} \\ 
\hline
\end{tabular}
\caption{Coefficients of the linear Higgs couplings in \eq{Leff}, for the SM, the $\SO(5)/\SO(4)$ composite Higgs (MCHM), and the dilaton Higgs.}
\label{hlinear}
\end{table}
In Table~\ref{hlinear} we have defined $\xi = v^2/f^2$, and notice first the important fact that in the MCHM the deviations from the SM scale with $\xi$, thus the SM limit is reproduced for $\xi \to 0$. This is a common feature of all the composite Higgs models except for the dilatonic Higgs, where instead the SM limit is recovered when $\xi \to 1$. For the dilaton however this is not the only requirement to reproduce the SM. The anomalous dimensions of the SM operators, which encode the explicit breaking of scale invariance from the SM fields, must also vanish. These are associated to the Yukawa coupling of the fermion $\psi = t,b,\tau$, $\gamma_\psi$, and to the gauge field strength tensors, $\gamma_{g_i} = (b_{UV}^{(i)} - b^{(i)}_{IR} )g_i^2 /(4\pi)^2$. Importantly, the interactions of the dilaton with massless gauge fields receives its  leading corrections from the trace anomaly, in contrast with the MCHM where these corrections arise only after integrating out light composite states, generically subleading and not included in Table~\ref{hlinear}.
Let us also note that for the MCHM, the numerical factor multiplying $\xi$ in the coupling to electroweak gauge bosons, $1/2$ when expanded in powers of $\xi$, is fixed by the $\SO(5)/\SO(4)$ symmetry. In larger cosets such factor might be different, for instance in $\SU(5)/\SO(5)$ it is $1/8$.
However, one should bear in mind that if the additional GB's in these extended cosets are decoupled via large explicit breakings, the prediction for $hVV$ should approach those of the MCHM (as long as custodial symmetry is preserved).\footnote{In the littlest Higgs model of \cite{ArkaniHamed:2002qy}, based on the $\SU(5)/\SO(5)$ coset, once the extra vector resonances are integrated out and the custodial breaking triplet VEV is fine-tuned to vanish, one obtains a factor $5/32$ \cite{Giudice:2007fh}. This is far from the MCHM, but only because the corrections $g_{SM}/g_*$ are important in that particular realization.}
Let us also point out that the Higgs interactions with fermions depend on the specific form of the fermion couplings to the composite sector, in particular on the embeddings into the global symmetries. Using the general structure presented in \cite{Pomarol:2012qf} for the mass of the fermion, $m_\psi(h) \propto \sin(h/f) \cos^{n_\psi}(h/f)$, with $m_W(h)=gf\sin(h/f)/2$, one can derive the $c_\psi$ presented in Table~\ref{hlinear}.

To parametrize this model dependence, the deviations in the Higgs couplings can be analyzed in general by encoding the effects of new-physics in higher-dimensional operators involving the Higgs complex doublet field \cite{Manohar:2006gz,Giudice:2007fh,Low:2009di}. The most relevant ones are: \emph{i)} Universal corrections to all Higgs couplings, arising as a modification of the Higgs kinetic term from the operator $(\partial_\mu (H^\dagger H))^2$. This is generically generated from the non-linear structure of the coset interactions, extra scalars mixing with the Higgs, tree-level exchange of vector partners, and at one loop from top partners and extra GB's. Notice that this term gives rise to modified Higgs coupling to electroweak gauge bosons correlated with the modification in the couplings to fermions; \emph{ii)} This correlation is broken by the operator $H^\dagger H \bar \psi_L H \psi_R$, which affects only the fermionic couplings of the Higgs; \emph{iii)} Given its importance for Higgs production and decay, the operators $H^\dagger H F_{\mu \nu}^2$ parametrize  the corrections of the Higgs couplings to massless gauge bosons. The contributions of these operators to the parameters of \eq{Leff} can be found in Table 1 of \cite{Contino:2013kra}.\footnote{In that table the contributions of several other operators to a more complete set of effective interactions of the Higgs are also shown, which are relevant for 3-body $V \psi \psi$ Higgs decays, $V = W, Z$.}
Several other works have also recently reassessed such effective Lagrangians in the context of the newly discovered Higgs boson~\cite{Alonso:2012px,Brivio:2013pma,Elias-Miro:2013mua,Pomarol:2013zra}. Given a proper complete basis of operators for physics beyond the SM, corrections and correlations on observables can be consistently derived, allowing for instance to identify which new physics Higgs signals are still poorly constrained \cite{Elias-Miro:2013mua,Pomarol:2013zra}. 
One particularly interesting unconstrained channel  is the $h \to Z \gamma$ decay rate~\cite{Elias-Miro:2013mua,Azatov:2013ura}.

The odd case is again that of the dilatonic Higgs~\cite{Chacko:2012vm,Bellazzini:2012vz}, where the proper effective Lagrangian disengages the longitudinal components of the $W$ and $Z$ from the Higgs particle, see for instance \cite{Alonso:2012px}.\\

\begin{figure}[t!]
\begin{center}
\includegraphics[width=3in]{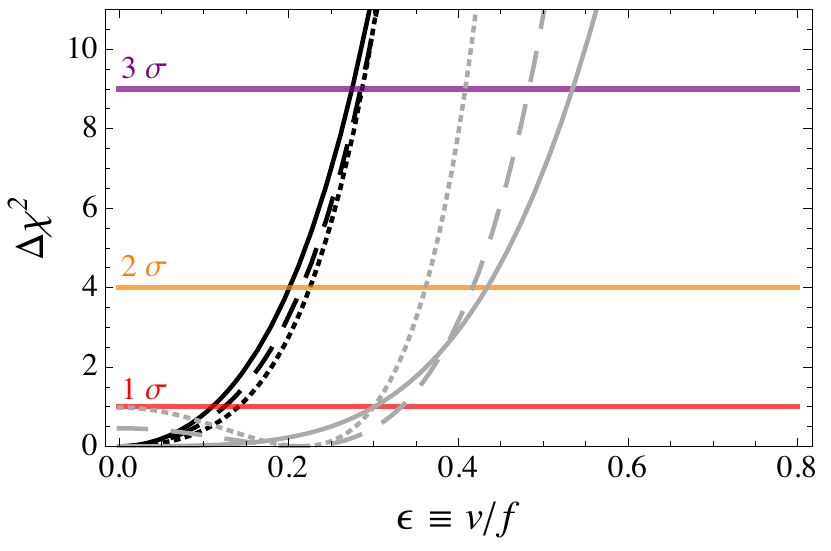}
\hspace{1cm}
\includegraphics[width=2.5in]{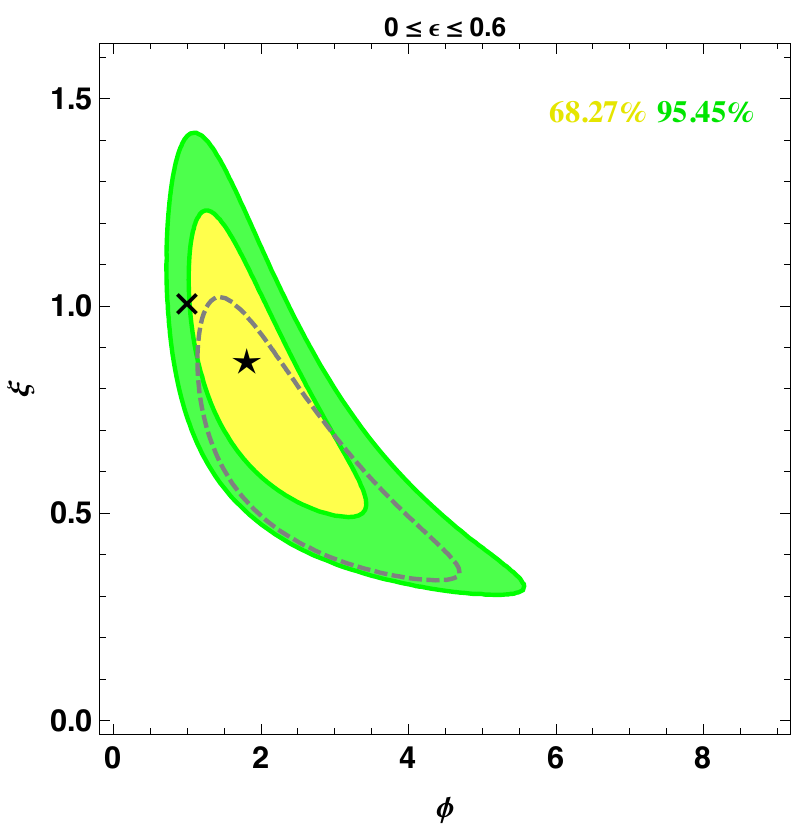}
\caption{Higgs fits from \cite{Falkowski:2013dza} (left panel) and \cite{Chacko:2012vm} (right panel). Left panel: Fit to $v/f$ for the MCHM with (black) or without (gray) including electroweak precision data, with $n_\psi = 0$ (solid), $n_\psi = 1$ (dashed), and $n_\psi = 2$ (dot-dashed). Right panel: Fit to $\xi = v^2/f^2$ and $c_{\gamma \gamma}/\xi$ from Higgs data, with $\epsilon \equiv \gamma_\psi$ marginalized in the range $0 \leqslant \epsilon \leqslant 0.6$. The star is the best-fit point, while the cross corresponds to Higgs-like dilaton limit.}
\label{fit}
\end{center}
\end{figure}

Given all these considerations, we come back to the particular models discussed above, to show in Fig.~\ref{fit} left panel the fit for the MCHM in terms of $v/f$ ($\epsilon$ in the plot) for $n_\psi = 0, 1, 2$, taken from \cite{Falkowski:2013dza}, and in the right panel for the dilaton in terms of $\xi = v^2/f^2$ and $c_{\gamma \gamma}/\xi$ ($\phi$ in the plot) for $\gamma_\psi$ ($\epsilon$ in the plot) between 0 and 0.6 and $c_{gg} = 0$, taken from \cite{Chacko:2012vm}.
For the MCHM, given the absence of significant deviations from the SM predictions, a lower bound on the compositeness scale $f \gtrsim 700 \GeV$ at 1$\sigma$ level can be obtained from Higgs couplings measurements only (gray lines), while $f \gtrsim 1.5 \TeV$ if the electroweak precision data, mostly affecting $c_V$, is included in the fit (black lines). As explained above, these bounds apply to most of the composite Higgs models, although they can be somewhat relaxed if there is an extended GB sector \cite{Montull:2012ik,Reuter:2012sd,Vecchi:2013bja} (see also \cite{Belanger:2013xza}), or extra contributions to $\hat T$ as explained in Sec.~\ref{ewpt}.
For the Higgs-like dilaton, if the electroweak precision data is not included there is still a significant allowed range for $\xi$ around 0.8, correlated with the values of $c_{\gamma \gamma}$ and $c_{gg}$, which in this case can display ${\cal O}(1)$ deviations from the SM. However, if the bound on $c_V$ from EWPT is taken into account, it forces $f \lesssim 300 \GeV$ and small anomalous dimensions $\gamma_{\psi}, \gamma_{g_i} \ll 1$.\\

At this point it is worth pointing out that deviations in the $h t \bar t$ coupling and direct contributions to the $h gg$ coupling both affect the main Higgs production channel via gluon fusion. Given that in models such as the MCHM, the leading new physics effects modify $c_t$, while for the dilaton it is $c_{gg}$ that receives the largest corrections, one important subject is to disentangle them. Several approaches have been proposed to achieve this: $t h$ production \cite{htproduction}, $t \bar t h$ production \cite{httproduction}, and $hj$ production \cite{hjproduction}.

It is thus clear that it is very important to identify which new physics contributions to the Higgs one-loop couplings to gluons and photons are predicted in scenarios such as the MCHM due to the presence of light states (the top and vector partners).  Since the Higgs is assumed to be a pGB, these contributions are expected to scale with $g_{SM}^2/m_{*}^2$, in addition to the loop factor $g^2/16\pi^2$.
Several analyses have considered such deviations due to the light top partners \cite{Low:2010mr,Gillioz:2012se,Farina:2013ssa}, and the same behavior is expected from the vector partners.
In these corrections collective symmetry breaking plays an important role: it basically eliminates the dependence on masses of the partners, leading to a shift in the $hgg$ coupling which scales as $v^2/f^2$, and independent of $g_{*}=m_{*}/f$ to leading order in $m_h^2/m_{*}^2$. This shift is the same as in the $h t \bar t$ coupling to leading order in $\lambda_{L,R}^2/g_{*}^2$.
This was first pointed out in~\cite{Falkowski:2007hz} in the holographic Higgs framework, followed by recent works~\cite{Azatov:2011qy,Montull:2013mla}.
This pattern of deviations no longer holds if the partners for the light SM fields, such as the bottom quark, are light. 
This is due to the fact that light SM fields do not contribute to loop-mediated Higgs couplings, while their partners do~\cite{Azatov:2011qy,Delaunay:2013iia}.
In this case the $h gg$ and $h \gamma \gamma$ couplings are  sensitive to the spectrum of partners, their corrections scaling with the expected $\lambda_{L,R}^2/m_{*}^2$ ratio.
Also, if the low-energy theory contains more than one LR Yukawa-type operator, non-trivial dependence on $m_{*}$ can arise.\footnote{Besides, Higgs plus jet production has been shown to display a higher sensitivity to the top partners masses and couplings \cite{Banfi:2013yoa}.} 
It is important to point out at this point that the GB suppression does not hold for $h Z \gamma$, since this coupling involves both a massless and a massive SM gauge boson, thus it  does not need to be suppressed by the GB symmetry~\cite{Azatov:2013ura}.
Finally, notice that in twin Higgs models, where the SM partners are not charged under the SM gauge symmetries, no effects are present except through Higgs operators~\cite{Craig:2013xia}.\\

The LHC and future hadron colliders expect to measure single Higgs couplings with a precision of a few percent, while future $e^+ e^{-}$ colliders such as CLIC, ILC and TLEP may reach sub-percent level for almost all couplings, see e.g. Table~1-20 in \cite{Dawson:2013bba} (based on original studies like \cite{Peskin:2012we,Klute:2013cx}). The measurement of a given Higgs coupling $c$ with precision $\delta_c$ allows one to set a $95\%$~CL bound on its deviations from the SM as large as $\sim 2\delta_c$. 
Alternatively, deviations larger than $\sim 5\delta_c$ allow one to claim a $5\sigma$ (indirect) discovery.  We can thus translate the expected precision on the measurement of a Higgs coupling to a expected bound (or reach for discovery) on composite Higgs models.
For example, in the MCHM one has $\delta c_V=1-c_V \simeq \xi/2$ which allows one to set the $95\%$ CL bound $\xi = v^2/f^2 < 4\delta_{c_V}$, or claim discovery for $\xi \gtrsim  10\delta_{c_V}$. Besides, combining information from the measurements of several couplings may also improve the bounds. 
Table~\ref{tab_boundxi} shows the bounds on $\xi$ and $f$ from the projected  precision on the measurements of $c_V$ at future colliders. The deviation in the coupling to fermions is more model dependent, $\delta c_\psi=(2n_\psi+1)\xi/2$, and  may give a stronger bound for the charged leptons $c_\ell$ and/or for $n_\psi > 1$, see Table~1-20 in \cite{Dawson:2013bba}.

\begin{table}
\tiny
\begin{center}
\hspace*{-0.6cm}
\begin{tabular}{lcccccccc}
\hline
Facility  &  LHC & HL-LHC & ILC500 & ILC500-up & ILC1000 &   ILC1000-up & CLIC  & TLEP (4 IPs) \\
$\sqrt{s}$ (GeV)  &  $14000$ & $14000$ & $250/500$ & $250/500$ & $250/500/1000$ & $250/500/1000$ & $350/1400/3000$ & $240/350$    \\
$\int\mathcal{L}dt$ (fb$^{-1}$)  &  $300$/expt & $3000$/expt & $250+500$ & $1150+1600$ & $250+500+1000$ & $1150+1600+2500$  & $500+1500+2000$  & $10000+2600$ \vspace{0.05cm} \\
\hline
\vspace{0.05cm} 
 $\xi = v^2/f^2$ & $0.16-0.24$ & $0.08-0.16$ & $0.016$ & $0.008$ & $0.008$ & $0.008$         & $0.02/0.006/0.004$ & $0.002$\\
\hline
\vspace{0.05cm}
$f$ (GeV) & $500-615$  &   $615-870$   &   $1950$  &   $2750$  & $2750$ & $2750$  & $1750/3175/3700$  & $5500$ \\
\hline
 \end{tabular}
 
 \caption{\label{tab_boundxi} Upper (lower) bounds on $\xi = v^2/f^2$  ($f$) at $95\%$ CL in the MCHM extracted from the expected precision on the Higgs coupling $c_V$ from the 7-parameter fit of \cite{Dawson:2013bba} (quoting the stronger bound from either $c_W$ or $c_Z$). 
The fit assumes no exotic production or decay modes, as well as generation universality for the coupling to fermions. The ranges shown for LHC and HL-LHC represent the conservative and optimistic scenarios where the systematic uncertainties are left unchanged or scaled with the increased statistics (on top of a $1/2$ factor for the theory uncertainties) respectively.  The lower bound on $\xi$ needed to reach a $5\sigma$ discovery can be estimated by multiplying  the quoted upper bounds by $5/2$.}
 \end{center}
 \end{table}

\subsubsection{Double-Higgs production}

We begin this section by noticing an important but obvious point. Since the recently discovered Higgs boson has SM-like couplings, in particular to the massive gauge bosons $V = W, Z$, the unitarization of their scattering amplitudes $VV \to VV$ is accomplished to a high degree by the Higgs itself, without the need of any new resonances up to at least $\sim 3 \TeV$ \cite{Contino:2010mh,Contino:2011np,Bellazzini:2012tv}.
For the case that the Higgs arises as a 4-plet of GB's, the above statement, in effective field theory language, is equivalent to the confirmation that the operator $(\partial_\mu (H^\dagger H))^2$ is suppressed by a scale $f$ hierarchically larger than the electroweak scale.
Furthermore, in this case the properties of the $W$ and $Z$ are intrinsically tied to those of the Higgs boson, and as such their behavior at high energies is completely correlated by the $\SO(4)$ symmetry. Because of this, the high energy behavior of double Higgs production does not offer a new (compared to $WW$ scattering) avenue where beyond the SM behavior might be expected. However, two important comments are in order. First, there is a composite Higgs candidate which does not exhibit the above features by construction: the dilatonic Higgs. Second, the production of Higgs boson pairs can be affected by several other new-physics effects, as we now show.

As in the previous section, we parametrize the double interactions of the Higgs by a phenomenological Lagrangian \cite{Contino:2012xk}
\begin{eqnarray}
\mathcal{L}_{eff}^{(h^2)} &=&
\left( \frac{d_V}{2} \left( m_W^2 W^{+}_\mu W^{-\mu} + m_Z^2 Z_\mu^2 \right) - d_t m_{t} \bar{t} t  - d_b m_{b} \bar{b} b - d_\tau m_{\tau} \bar{\tau} \tau \right) \frac{h^2}{v^2} \nn \\
&& + \left( \frac{d_{gg} }{2} G^{a}_{\mu \nu} G^{a,\mu \nu}\right) \frac{h^2}{v^2}  - \frac{c_3 }{2} \frac{m_h^2}{v} h^3
\, ,
\label{Lh2eff}
\end{eqnarray}
and present in Table~\ref{hdouble} the predictions for the MCHM of \cite{Agashe:2004rs}, and the dilatonic Higgs \cite{Bellazzini:2012vz}. For the MCHM we omit again the effects of the light SM partners, but we comment on those below. As for the linear Higgs couplings, the deviations from the SM vanish in the limit $\xi \to 0$ for the MCHM,  as well as in other models where the Higgs boson belongs to the same multiplet as the scalars eaten by the $W$ and the $Z$.  Once again, the dilaton mimics the SM prediction in the opposite limit $\xi \to 1$, along with vanishing anomalous dimensions, except for one notable exception, the trilinear Higgs self-interaction $c_3$. 
\begin{table}[t!]
\centering
\begin{tabular}{cccc}
\hline
coupling & SM & MCHM & Dilaton \\ \hline
$d_V$ & 1 & $1-2\xi$ & $\xi$ \vspace{0.1cm} \\ 
$d_\psi$ & 0 & $\frac{-\xi(1+3 n_\psi-(1+n_\psi)^2 \xi)}{2(1-\xi)}$ &
$\frac{1}{2} \gamma_\psi \xi$ \vspace{0.1cm} \\ 
$d_{gg}$ & 0 
& 0 \vspace{0.1cm} & 
$-\frac{\alpha_s}{8 \pi} ( b_{IR}^{(3)}-b_{UV}^{(3)} ) \xi$ \\ 
$c_3$ & 1 & $\frac{1-(1+\tilde n_\psi) \xi}{\sqrt{1-\xi}}$ & 
$\frac{1}{3} (5+d\beta/d\lambda) \sqrt \xi$ \\ 
\hline
\end{tabular}
\caption{Higgs couplings in \eq{Lh2eff} for the SM, the MCHM, and the dilaton.}
\label{hdouble}
\end{table}
This can be understood by noticing that the SM result $c_3 =1$ is reproduced if the perturbation explicitly breaking scale invariance is a pure mass term, as in the SM, since then $d\beta/d\lambda = -2$ (where in the SM case $\lambda = \mu$).
However, the natural realization of the Higgs-like dilaton scenario (with a sufficiently light dilaton) implies $d\beta/d\lambda \propto m_d^2/\Lambda_{C}^2$, which makes this a subleading contribution. This fact then establishes double-Higgs production as the key test 
to distinguish the dilatonic Higgs scenario from an ordinary Higgs.
Let us also note that for the MCHM the numerical factor multiplying $\xi$ in $d_V$ is again fixed by the $\SO(5)/\SO(4)$ symmetry, and for larger cosets these coefficients could be different. This also applies to double Higgs couplings to fermions, which are embedding dependent, and which we have derived again from $m_\psi(h) \propto \sin(h/f) \cos^{n_\psi}(h/f)$.
The prediction for $c_3$ in the MCHM is more model dependent, since it depends on what the leading contribution to the Higgs potential is. We have assumed here that it is of the form $V(h) = \cos^{1+\tilde n_\psi}(h/f) \left(\alpha - \beta \cos^{1+\tilde n_\psi}(h/f) \right)$.
All this model dependence can again be encoded in the coefficients of higher-dimensional operators beyond the SM, in particular $(\partial_\mu (H^\dagger H))^2$, $H^\dagger H \bar \psi_L H \psi_R$, $H^\dagger H G_{\mu \nu}^2$, and $(H^\dagger H)^6$, for $d_V$, $d_\psi$, $d_{gg}$, and $c_3$ respectively~\cite{Contino:2010mh}.
In any case it is important to stress that double-Higgs production via gluon fusion is not only sensitive to the trilinear Higgs coupling, but also to the $hh t \bar t$ and $hh gg$ couplings. The actual sensitivity is more promising for the latter rather than the former \cite{Contino:2012xk}.
The effects of the top partners on these couplings have also been studied \cite{Gillioz:2012se}, with the important result that the process $gg \to hh$ gets sizable contributions, contrary to the expectations for single-Higgs production in $gg \to h$.\\

Let us conclude this section with more comments on the high energy behavior of $W,Z$ and $h$ scattering. We stress the fact that in most composite Higgs models at the high energies the relation  $\mathcal{A}(W^+W^- \!\to\! hh) \simeq \mathcal{A}(W^+W^- \!\to\! ZZ)$ is expected to hold due to the Higgs being part of an $\SO(4)$ vector, unlike for the dilaton.  The relation between the linear and double dilaton couplings to the massive gauge bosons $V = W, Z$ ensures that the growth with energy in $VV \to hh$ is absent at leading order $\mathcal{A}(W^+W^- \to hh) \simeq (d_V-c_V^2) (s/v^2) = 0$.  However this relation is affected by higher derivative terms, such as $\partial_\mu h \partial_\nu h \partial^\mu \partial^\nu h$ or $2 m_V^2 V_\mu V_\nu \partial^\mu h \partial^\nu h$. The first of these operators breaks the $h \to - h$ parity symmetry present in the chiral Lagrangian of the MCHM (a property that is shared by all the composite Higgses except for the dilaton). The feasibility of probing these interactions at the LHC is quite limited, with better perspective at a linear collider \cite{Contino:2013gna}.

\subsubsection{Invisible decays}
Composite Higgs models providing a dark matter candidate may predict invisible Higgs decays which in turn affect the various  branching ratios into visible final states. Because of the small Higgs width in the SM, $\Gamma_{SM}\sim 10^{-5} m_h$, even relatively small couplings of the Higgs boson to dark matter (or to other undetectable final states) may result into relatively large modifications of the branching ratios. CMS has placed a direct upper bound of $69\%$ (at $95\%$ CL) on the invisible branching ratio in the VBF channel \cite{CMSinvisibleVBF}. The upper bounds on the Higgs invisible branching ratio in the $Zh$ associated production channel are $75\%$ from CMS \cite{CMS:2013yda}, and $65\%$ from  ATLAS \cite{ATLAS:2013pma}. The invisible Higgs branching ratio is also constrained indirectly by $BR_{inv}\lesssim 0.6$ \cite{ATLAS:2013sla,Espinosa:2012vu} obtained from fitting the Higgs couplings. Stronger bounds in the $35-50\%$ range can be obtained by allowing variations of the Higgs couplings to gluons and photons in the fit \cite{Giardino:2013bma,Falkowski:2013dza}.

\subsection{Direct searches} \label{resonances}

The SM partners are constrained indirectly from electroweak, flavor, and Higgs physics, as we have reviewed in the previous sections. Already from LEP the bounds on generic vector partners were fairly strong, $m_\rho \gtrsim 2.5 \TeV$. On the other hand pre-LHC bounds on fermion partners were less constraining, and LHC Higgs couplings measurements are not contributing much to the bounds on the partner masses. Nevertheless, these indirect measurement can be sensitive to the UV properties of the models around the strong coupling scale $\Lambda_C$, while direct searches do not have that problem. The latter thus constitute a direct probe of the fine-tuning in any given model.

There are many studies on the phenomenology of the SM partners, either in little Higgs \cite{ArkaniHamed:2002pa,Hewett:2002px,Han:2003wu,Han:2005ru} or in holographic Higgs models \cite{Carena:2006bn,Carena:2007tn}. We will classify them based on the spin.\\

{\bf Spin-1 gauge partners}: These vector resonances are the  $W_H,Z_H$ gauge boson partners in little Higgs models \cite{Burdman:2002ns,Conley:2005et}, in warped extra dimensions they are the KK gauge bosons~\cite{Agashe:2007ki,Agashe:2008jb,Agashe:2009bb}, or generically they are simply $\rho$ mesons. These states could have played an important role~\cite{Bellazzini:2012tv} in the unitarization of the $WW$ scattering amplitudes, however since the Higgs couplings are  SM-like there is not much need for that. Therefore their main role is to tame the radiative contributions to the Higgs potential from the $\SU(2)_L \times \U(1)_Y$ gauge bosons. For studies of the 4D general effective Lagrangians describing these fields see~\cite{delAguila:2010mx,Contino:2011np,Bellazzini:2012tv}.

Due to the strong indirect bounds, we focus on the limit of strong coupling $g_\rho\gg g$ (which increases the mass of the $\rho$'s to several TeV). These resonances have coupling $g_\rho$ to the composite states (including the Higgs and longitudinal gauge bosons), while the coupling to quarks, leptons and transverse gauge boson are expected to be significantly smaller, $g^2/g_\rho$ (unless one has a $\U(3)$ flavor symmetry \cite{Redi:2013eaa} and light quark compositeness, or simply $g_\rho \sim g$, though the latter is disfavored by EWPT's). Note that these latter couplings are not necessary to cut off the Higgs potential. In this case the branching ratios of the $\rho$'s are dominated by the decays $\rho \to WW, WZ, Wh, Zh$. Also decays to $t \bar t, t \bar b$  are plausible given the assumption of the compositeness of the top. Moreover, given the necessary hierarchy implied by the constraints and the fine-tuning arguments, decays to top partners could actually dominate. The production of the $\rho$'s is expected to be dominated by single Drell-Yan production, through their mixing with the $W$ and $Z$. Another important channel might be associated production with jets if they are coupled more strongly to light quarks. At a linear collider, effects on $e^+ e^- \to f \bar f$ due to the $\rho$'s have been studied for instance in \cite{Conley:2005et}.\\

While 4D models do not necessarily include them, excitations of the gluon are an integral part of most extra dimensional models, and have been thoroughly investigated~\cite{Agashe:2006hk,Lillie:2007yh,Lillie:2007ve}. In fact this is one of the most prominent signals of the extra dimensional versions, due to the enhanced production rate of the KK gluons at hadron colliders. It is plausible that such color-octet excited states show up in generic models as well, since some of the fields in  the composite sector must be charged under color in order to be able to generate the top partners (even though the mass of the gluon partners has no direct connection with naturalness).\\

\begin{figure}[t!]
\begin{center}
\includegraphics[width=2.5in]{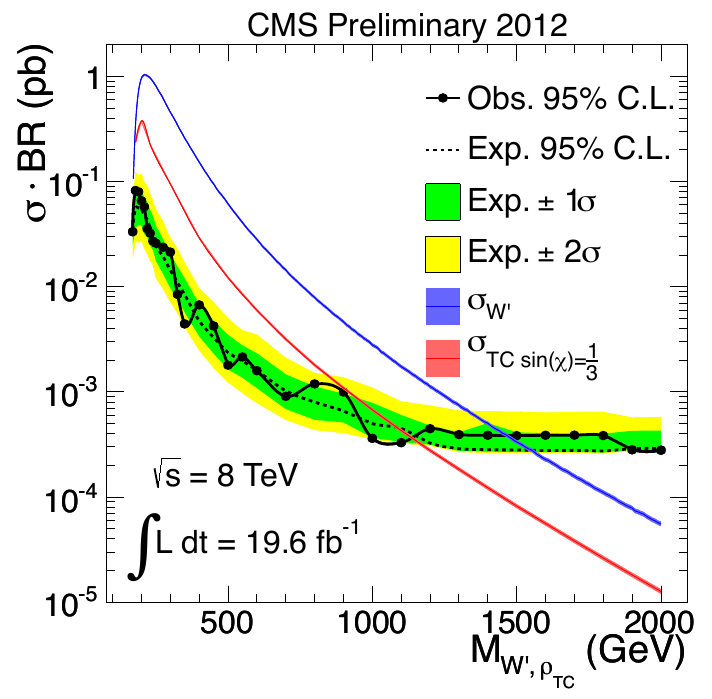}
\hspace{0.2cm}
\includegraphics[width=3in]{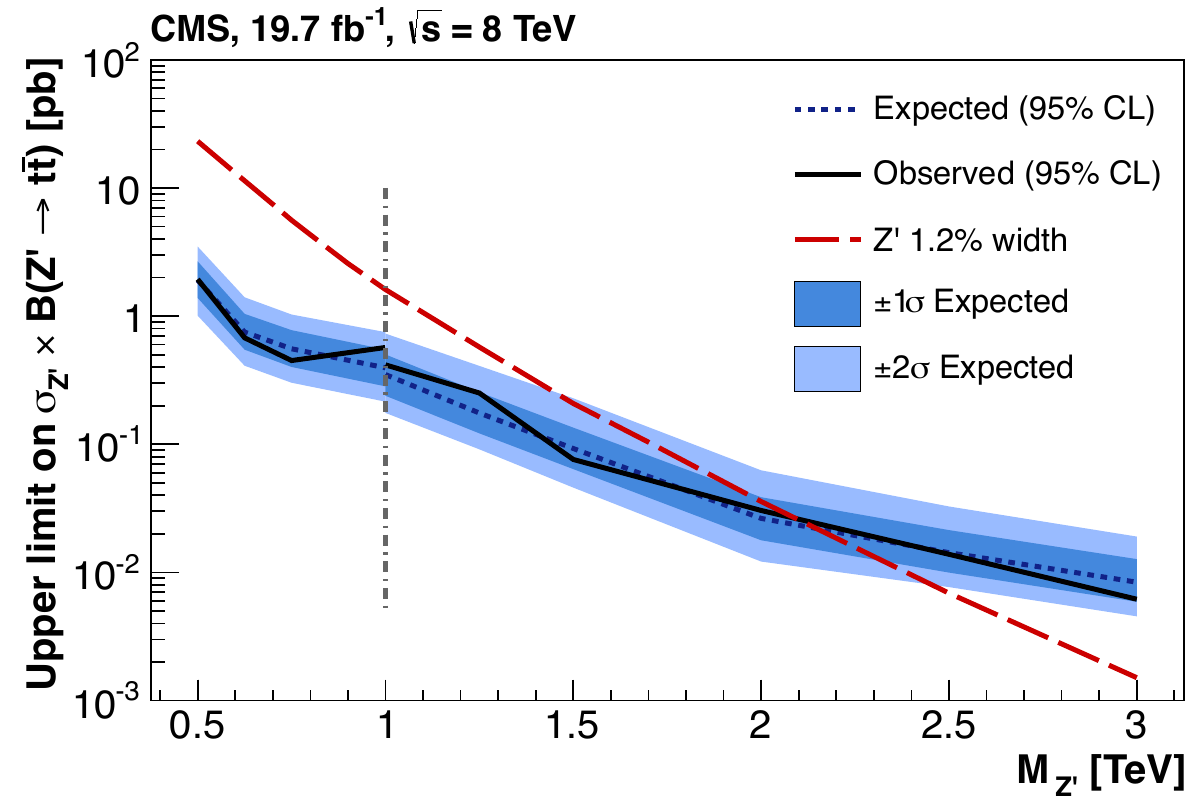}
\caption{Preliminary CMS bounds from run 1 of the LHC on the production of spin 1 resonances. Left panel: bound on $\rho^\pm$ using decays to $WZ$, from~\cite{CMS:2013vda}. Right panel: bound on the KK gluon decaying to $t \bar t$, from~\cite{Chatrchyan:2013lca}. Note that the dashed curve is for a $Z'$, the KK gluon bound from the same plot is around 2.5 TeV. }
\label{rho}
\end{center}
\end{figure}

The direct searches at ATLAS and CMS are most sensitive to $\rho^\pm$ production with decays to $WZ$. The final CMS run 1 bound is $m_\rho \gtrsim 1.1 \TeV$ at 95\% CL  ($\sim 20 \fb^{-1}$ at 8 TeV) \cite{CMS:2013vda}, see Fig.~\ref{rho} left panel. One obtains similar bounds in ATLAS \cite{ATLAS:2013lma} although the integrated luminosity in the most recent analysis is somewhat lower $\sim 14 \fb^{-1}$, leading to a slightly reduced bound.
Important constraints can arise also from resonance searches in $t \bar t$ production. The resulting bounds  depend on the degree of compositeness of the top, generically for the $\rho$ they are milder  than those from $WZ$ searches. On the other hand for the KK gluon this is the leading channel, since the branching ratio is usually strongly dominated by $t\bar t$. The resulting run 1 CMS bound is $m_{G} \gtrsim 2.5 \TeV$ at 95\% CL ($\sim 20 \fb^{-1}$ at 8 TeV) \cite{Chatrchyan:2013lca} (and again slightly weaker for ATLAS due to less luminosity \cite{ATLAS:2013kha}). Notice that if the decays to $t \bar t$ and $t \bar b$ are non-negligible then the $BR$ to $VV$ and $Vh$ will be reduced, thus the above bounds can be weakened  (to date no analysis for a combined bound in both channels has been performed). \\

{\bf Spin-1/2 top partners}: The investigation of the phenomenology and collider physics of the top partners has been initiated in the framework of the Little Higgs models~\cite{Perelstein:2003wd}, for  recent analyses in this context see~\cite{Berger:2012ec,Godfrey:2012tf}. As discussed throughout this review, these states are also predicted in the warped extra-dimensional models or pure 4D descriptions~\cite{Contino:2008hi,Mrazek:2009yu,Dissertori:2010ug,Vignaroli:2012nf,Cacciapaglia:2012dd,Li:2013xba} as  they are responsible for taming the radiative contributions to the Higgs potential from the top quark. Recent analyses of 4D effective Lagrangian descriptions parametrizing the most general possible interactions of the top partners can be found in~\cite{AguilarSaavedra:2009es,DeSimone:2012fs,Buchkremer:2013bha,Aguilar-Saavedra:2013qpa}.

The properties of the top partners depend on their quantum numbers under the global symmetries of the composite sector. If custodial $\SO(4)$ is assumed, it is common to find a $\mathbf{4}$ (required to couple to $q_L$) and $\mathbf{1}$ (to coupled to $t_R$). In almost all composite models they are triplets of color (the exception being twin Higgs models). Searches are typically classified by their electric charges: $T_{5/3}$, $T_{2/3}$, and $T_{-1/3} \equiv B$, although even more exotic charges have been proposed e.g. $T_{8/3}$ \cite{Pappadopulo:2013vca}, arising from a $\mathbf{9}$ of $\SO(4)$.

The phenomenology of the top partners depends on their production and decay. The leading gluon fusion initiated production is more model independent. However single production via $W,Z$ exchange is also very important for relatively heavy states. 
Their decays are usually fixed by symmetry. The Goldstone boson equivalence theorem mostly fixes the couplings and therefore the decay rates:
\emph{i)} $BR(T \to Zt) \simeq BR(T \to ht) \simeq BR(T \to W^+b)/2$ for the $T_{2/3}$ singlet under $\SO(4)$ (or $\SU(2)_L$).
\emph{ii)} $BR(B \to W^-t) \simeq 1$ for the $B$ doublet (under $\SU(2)_L$).
\emph{iii)} $BR(T \to Zt) \simeq BR(T \to ht)$ for the $T_{2/3}$ doublet.
\emph{iv)} $BR(T \to W^+t) \simeq 1$ for the $T_{5/3}$ doublet.
It is important to recall that this is somewhat dependent on the spectrum. There could be cascade decays or extra light GB's that can reduce the branching ratios~\cite{Kearney:2013oia,Kearney:2013cca}. The phenomenology of  composite light generations with various flavor symmetries can be found in~\cite{Delaunay:2013pwa}.\\

\begin{figure}[t!]
\begin{center}
\includegraphics[width=1.9in]{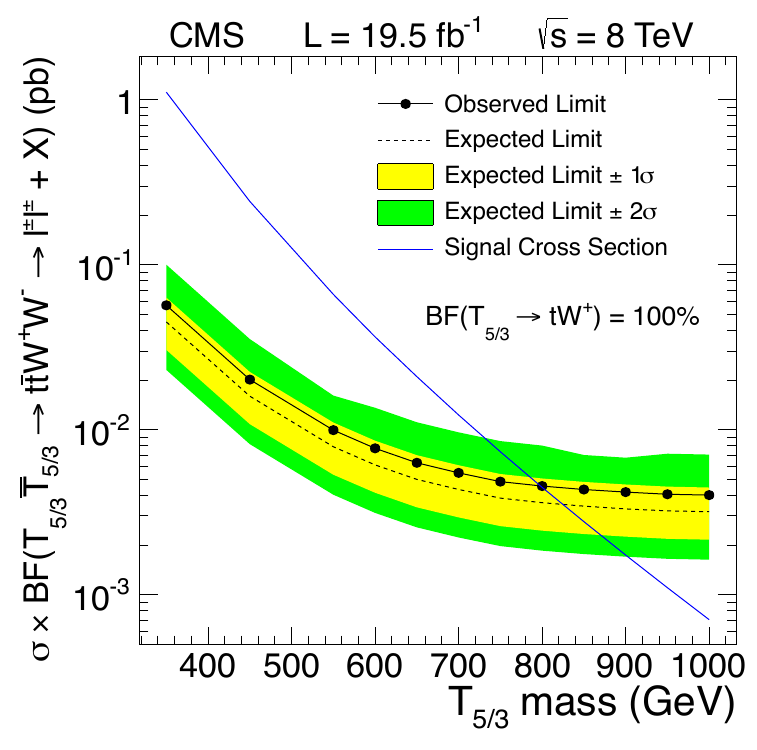}
\hspace{0.1cm}
\includegraphics[width=2.2in]{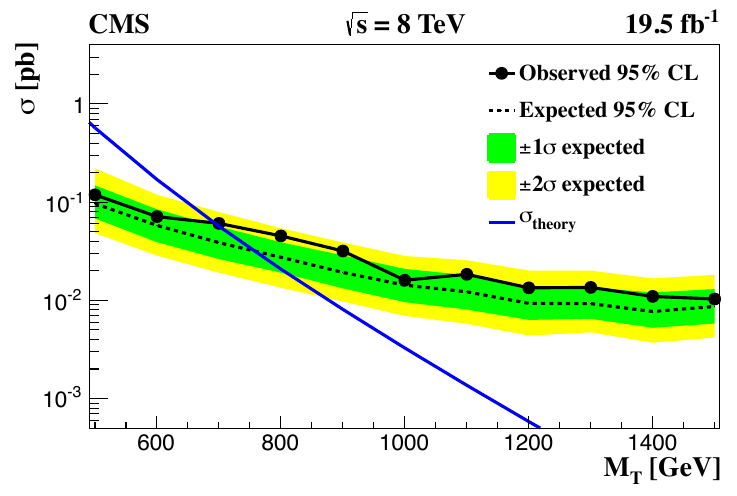}
\hspace{0.1cm}
\includegraphics[width=2.2in]{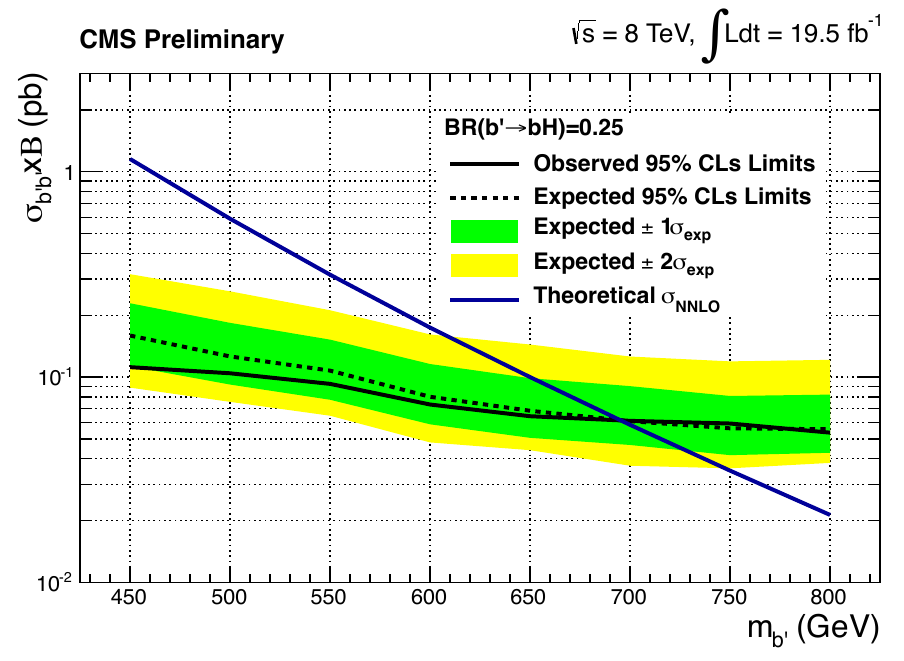}
\caption{CMS bounds on spin 1/2 top partners after run 1. Left: the bound on the charge $5/3$ top partner from~\cite{Chatrchyan:2013wfa}, middle: the bound on the charge $2/3$ singlet from~\cite{Chatrchyan:2013uxa}, right: the bound on the charge $-1/3$ singlet bottom partner from~\cite{CMS:2013una}.}
\label{topCMS}
\end{center}
\end{figure}

The  95\% CL final run 1 bounds from CMS using $\sim 20 \fb^{-1}$ luminosity at 8 TeV are shown in Fig.~\ref{topCMS}: $m_{T_{5/3}} \gtrsim 800 \GeV$ left panel \cite{Chatrchyan:2013wfa}, $m_{T_{2/3}} \gtrsim 700 \GeV$ for the singlet, middle panel \cite{Chatrchyan:2013uxa}, $m_{\tilde{B}} \gtrsim 700 \GeV$ right panel \cite{CMS:2013una}, where $\tilde{B}$ is singlet under $\SU(2)_L$, thus $BR(\tilde B \to Zb) \simeq BR(\tilde B \to hb) \simeq BR(\tilde B \to W^-t)/2$. The references also contain limits for ``non-standard'' branching ratios.

The most recent ATLAS bounds are for slightly lower luminosity $\sim 14 \fb^{-1}$ at 8 TeV, yielding somewhat milder bounds. The ATLAS analyses are organized by collider signatures and thus they apply to several top partners: lepton plus jets corresponding to mainly $T_{2/3} \to h t$ is found in~\cite{ATLAS:2013ima}, and $T_{2/3} \to W^+ b$ is in~\cite{ATLAS:2013sha}, same-sign dileptons corresponding to all possible kinds of $T_{2/3}$ and $B$ decays (singlets and doublets under $\SU(2)_L$)\cite{ATLAS:2013jha}, and $Z$ plus jets corresponding mainly to $T_{2/3} \to Zt$ and $B \to Zb$ decays (singlets as well as doublets) can be found in~\cite{ATLAS:2013oha}.
These analyses also provide limits as a function of branching ratios.\\

Future hadron colliders are expected to significantly improve the sensitivity to the various spin 1/2 partners. In Table~\ref{tab:futurespinhalf} we have summarize the estimated sensitivities as collected in Table 1-15 of~\cite{SnowmassTop} (based on the analysis of \cite{futuretop}) for the 14 TeV LHC and a 33 TeV upgrade of it. Studies for bounds on top partners at 100 TeV hadron colliders are currently being performed. 
 
\begin{table}
\begin{center}
\begin{tabular}{c|c|c|c} 
& LHC14 300 fb$^{-1}$ & LHC14 3000 fb$^{-1}$ & LHC 33 TeV 3000 fb$^{-1}$ \\ \hline
$T_{2/3}$ & 1.2 TeV & 1.45 TeV & 2.4 TeV \\
$B_{1/3}$ & 1.08 TeV & 1.33 TeV & $>$ 1.5 TeV \\
$T_{5/3}$ & 1.39 TeV & 1.55 TeV & 2.35 TeV \\
\end{tabular}
\end{center}
\caption{\label{tab:futurespinhalf} 5$\sigma$ discovery reaches of future hadron machines for various spin 1/2 partners, using pair production, from Table 1-15 of~\cite{SnowmassTop}.}
\end{table}

\subsection{Dark matter}

Dark matter (DM) candidates in composite models are of several nature. 
We can find partners of the SM states that enjoy a protecting global or discrete symmetry that renders them stable.  Alternatively, the coset space $\mathcal{G}/\mathcal{H}$ may have non-trivial homotopy groups and give rise to topologically conserved charges.

Non-trivial homotopy groups $\pi_n(\mathcal{G}/\mathcal{H})$ lead to $(2-n)$-dimensional defects such as domain walls ($n=0$), strings ($n=1$), and magnetic monopoles ($n=2$) whose cosmological abundances were studied e.g. in \cite{Murayama:2009nj}. The case of Skyrmions, $\pi_3(\mathcal{G}/\mathcal{H})\neq 0$,  has been explored recently within Little Higgs models in \cite{Joseph:2009bq,Gillioz:2010mr} where it was shown that the geometric annihilation cross-section $\sigma=\pi\langle r^2 \rangle$  may account for the observed DM relic density provided a quite large Skyrme parameter  is chosen. One generic problem of the models based on skyrmions is the stability of their masses and sizes which is achieved by balancing two operators with different dimensions, going beyond the regime of validity of the EFT. Nevertheless, there  exist 5D realizations \cite{Pomarol:2007kr} where the size of the skyrmion is in fact larger than the inverse cutoff of the theory and the predictions can thus be trusted. 

Models with extra conserved $\U(1)$'s were proposed originally within technicolor models \cite{Nussinov:1985xr,BarrChivukulaFarhi} where the lighest ``technibaryon'' (which may or may not be a PNGB) is stable and can have the observed DM relic density \cite{Gudnason:2006ug,Gudnason:2006yj,Foadi:2008qv} which is typically linked to the ordinary baryon asymmetry similarly in asymmetric DM models \cite{Kaplan:2009ag,Barbieri:2010mn}.

Other models with conserved $\U(1)$ baryon and lepton numbers have been considered within holographic versions of composite grand unified theories \cite{Frigerio:2011zg,Agashe:2002pr,Agashe:2004ci} where the $U(1)$'s are gauged and then spontaneously broken at the UV brane. Similarly to $R$ parity in SUSY, the resulting accidental $\textrm{Z}_n$ symmetry is enough to ensure DM stability over cosmological time scales \cite{Vecchi:2013xra}. 

Models with large cosets may give stable PNGBs by invoking suitable discrete symmetries acting on $\mathcal{G}/\mathcal{H}$.
For example, the next-to-minimal composite Higgs model $\textrm{O}(6)/\textrm{O}(5)$ studied in  \cite{Frigerio:2012uc} features an extra PNGB $\eta$ that is a SM singlet stabilized by one of the $\textrm{O}(6)$ parities, $\eta\rightarrow -\eta$. Interestingly, the model is particularly predictive in the region of parameter space that is consistent with the latest bounds from the LUX \cite{Akerib:2013tjd} and XENON100 \cite{Aprile:2012nq} experiments.  In particular, the $\eta$ can provide all the relic DM abundance, while naturally accommodating all the constraints, by choosing $m_\eta\gtrsim 100$~GeV and $f\sim1$~TeV. In this case, the annihilation cross-section mediated by the Higgs boson is controlled only by $f$ which fixes all the PNGB derivative coupling terms of the states parametrizing the coset as
\begin{equation}
\mathcal{L}=\frac{1}{2}(\partial_\mu\eta)^2+\frac{1}{2f^2}\left(\partial_\mu|H|^2+\frac{1}{2}\partial_\mu\eta^2\right)^2 \ .
\end{equation}
Notice also that in the regime $m_\eta < m_h/2$ bounds from the invisible $BR$ of the Higgs boson are among the strongest in this scenario \cite{Frigerio:2012uc}.

Models with T-parity~\cite{Cheng:2003ju,Cheng:2004yc} naturally contain a dark matter candidate, the lightest T-odd particle. Within little Higgs models this often turns out to be the partner of the neutral gauge boson $B$. A lot of work has been devoted to analyzing the viability of this scenario~\cite{Hubisz:2004ft,TparityDM}.

\section{UV completions\label{sec:UVcompletions}}
\setcounter{equation}{0}
\setcounter{footnote}{0}

The models presented here are all effective theories with a cutoff scale $\Lambda_C \simeq 4 \pi f \sim 5-10$ TeV. An important question is what these theories would look at a scale beyond the cutoff, which is not too far above the LHC energies. This motivates the search for UV completions. Assuming that one wants to avoid reintroducing the hierarchy problem, UV completions generically fall into two categories. The first are non-supersymmetric strongly coupled theories similar to QCD/Technicolor, but with modified dynamics. In this case one needs to guess the right symmetry breaking pattern and low-energy degrees of freedom, which should then be verified by lattice simulations. The second are supersymmetric UV completions, which may also involve some strong dynamics (but is usually under control due to the added constraint of supersymmetry).

One should emphasize that there are several different ways of trying to combine the pGB Higgs ideas with supersymmetry. In many cases, the low-energy theory (at a few 100 GeV) is actually a SUSY theory, which due to the pGB nature of the Higgs has interesting properties different from the ordinary MSSM. These include the so-called super-little Higgs~\cite{superlittle} and buried Higgs~\cite{buried} models. A particularly interesting SUSY model is that where only the idea 
of partial compositeness is implemented~\cite{partiallycompositeSUSY} $-$ due to SUSY there is no need to further protect the Higgs potential. Partial compositeness could rather raise the physical Higgs mass, and also provide a reason for hierarchical soft breaking terms~\cite{lightstop}. Purely composite SUSY Higgs models usually go under the name of ``fat Higgs"~\cite{fatHiggs}.
While all of these models contain some of the ingredients used in the non-SUSY pGB composite Higgs models, they are not true UV completions, since there is no regime where the theory is truly a non-supersymmetric composite Higgs model, with only a composite Higgs, the top partners and the vector partners in the spectrum. An attempt at such a SUSY UV completion for the MCHM was recently proposed in~\cite{Caracciolo:2012je}: the effective theory below 10 TeV is the SO(5)/SO(4) MCHM with top and vector partners (and perhaps a few scalar superpartners of the top partners). Other superpartners show up at 10 TeV. The model is based on the SO(4)$_m$ magnetic dual of a strongly coupled electric SO(N) theory, where the flavor symmetries contain an additional SO(5) factor. A different type of SUSY UV completion is based on a weakly coupled SUSY theory, a concrete example has been worked out for the case of little Higgs models in~\cite{Csaki:2008se,Pappadopulo:2010jx}.

The non-supersymmetric UV completions include a strongly coupled (non-QCD-like) SO(7) theory for the littlest Higgs model~\cite{Katz:2003sn}, as well condensing 4-Fermi operators a la NJL~\cite{Barnard:2013zea}.

Of course many of the composite Higgs models originate from extra-dimensional constructions. These have their own cutoff scales, which depends on the parameters of the theory. The theory below the cutoff generically describes the first few weakly coupled KK modes of the theory, the lightest of which can be identified with the top and gauge partners. However, to find a true UV completion one either needs to find a string theory construction, or use a deconstructed version without elementary scalars.

\section*{Acknowledgements}

We thank Ignatios Antoniadis and Dumitru Ghilencea for tasking us with this review, and for their patience with us.  We also thank Maxim Perelstein for useful conversations. B.B. thanks Filippo Sala for reading and commenting on the paper. B.B. is supported in part by the MIUR-FIRB grant RBFR12H1MW, and by the Agence Nationale de la Recherche under contract ANR 2010 BLANC 0413 01. C.C. and J.S. are supported in part by the NSF grant PHY-0757868.






\end{document}